\documentclass[aps,prd,reprint,showpacs,
groupedaddress,
nobibnotes,nofootinbib]{revtex4}
\usepackage{amsmath,amssymb,amsthm}
\usepackage{graphicx}
\usepackage{mathrsfs}
\usepackage{multirow}

\newcommand{\beq}{\begin{equation}}
\newcommand{\eeq}{\end{equation}}
\newcommand{\bw}{\begin{widetext}}
\newcommand{\ew}{\end{widetext}}

\begin{document}
\title{The Deformation of Poincar\'e Subgroups Concerning Very Special Relativity}

\author{Lei Zhang}

\author{Xun Xue}
\email[Corresponding author: ]{xxue@phy.ecnu.edu.cn}

\affiliation{Institute of Theoretical Physics, Department of Physics, East China Normal University, No.500, Dongchuan Road, Shanghai 200241, China}

\affiliation{Kavli Institute for Theoretical Physics China at the Chinese Academy of Sciences,Beijing, China}

\begin{abstract}
We investigate here various kinds of semi-product subgroups of Poincar\'e group in the scheme of Cohen-Glashow's very special relativity along the deformation approach by Gibbons- Gomis-Pope.  For each proper Poincar\'e subgroup which is a semi-product of proper lorentz group with the spacetime translation group $T(4)$, we investigate all possible deformations and obtain all the possible natural representations which inherit from the $5-d$ representation of Poincar\'e group. We find from the obtained natural representation that rotation operation may have additional accompanied scale transformation in the case of the original Lorentz subgroup is deformed and the boost operation get the additional accompanied scale transformation in all the deformation cases. The additional accompanied scale transformation has strong constrain on the possible invariant metric function of the corresponding geometry and the field theories in the spacetime with the corresponding geometry.
\end{abstract}

\maketitle

\section{Introduction}

The local Lorentz symmetry and CPT invariance is one of the fundamentals of modern physics. The theoretical investigation and experimental examination of Lorentz symmetry have made considerable progress and attracted a lot of attentions since the mid of 1990s. It is inevitable to encounter quantum gravity in the exploration of the theoretical framework of high energy physics, especially around the energy scale near Planck scale. Different quantum gravity models neither exclude Lorentz violation nor predict it conclusively. There are some high energy models of spacetime structure, such as non-commutative field theory, do however explicitly contain Lorentz violation. So the possible Lorentz violation is an important theoretical question \cite{Mattingly:2005re}.

There are many attempts to investigate the possible Lorentz violation from theoretical aspect\cite{Coleman:1998ti,Kosteleck:1998lv,Amelino-Camelia:2000mn, Amelino-Camelia:2010di,Girelli:2007pg}.
Because at low energy scales, parity $P$, charge conjugation $C$ and time reversal $T$ are individually good symmetries of nature while there is evidence of $CP$ violation for higher energies, one may consider the possible failure of Poincar\'e symmetry at such high energy scales. One theoretical possibility is that the spacetime symmetry of all the observed physical phenomena might be some proper subgroups of the Lorentz group along with the spacetime translations only if these kind of proper subgroups of Poincar\'e group incorporating with either of the discrete operations $P$, $T$ $CP$ or $CT$, can be enlarged to the full Poincar\'e group. The Very Special Relativity (VSR) proposal by Cohen and Glashow is based on these smaller subgroups \cite{Cohen:2006vy}.  Cohen and Glashow argued that the local symmetry of physics might not need to be as large as Lorentz group but its proper subgroup, while the full symmetry restores to Poincar\'e group when discrete symmetry $P$, $T$ $CP$ or $CT$ enters. The Lorentz violation is thus connected with CP violation. Since CP violating effects in nature are small, it is possible that Lorentz-violating effects may be similarly small. They identified these VSR subgroups up to isomorphism as T(2) (2-dimensional translations) with generators $T_{1}=K_{x}+J_{y}$ and $T_{2} = K_{y} - J_{x}$, where $\mathbf{J}$ and $\mathbf{K}$ are the generators of rotations and boosts respectively, E(2) (3-parameter Euclidean motion) with generators $T_{1},T_{2}$ and $J_z$, HOM(2) (3-parameter orientation preserving transformations) with generators $T_{1},T_{2}$ and $K_z$ and SIM(2) (4-parameter similitude group)with generators $T_{1},T_{2}, J_z$ and $K_z$. The semi-direct product of the SIM(2) group with the spacetime translation group gives a 8-dimensional subgroup of the Poincar\'e group called ISIM(2). The spurion strategy can also be applied to VSP. The invariant tensor for group $E(2)$ can be a 4-vector $n=(1,0,0,1)$ while the symmetry groups $T(2)$ admits many invariant tensors. There is neither invariant tensors for $HOM(2)$ and $SIM(2)$ nor the local Lorentz symmetry breaking perturbative discription for either of these groups.

Inspiring by the fact that Poincar\'e group admits the unique deformation into de Sitter group, Gibbons, Gomis  and Pope find that the subgroup ISIM(2) considered by Cohen and Glashow admits a 2-parameter family of continuous deformations which may be viewed as a quantum corrections or the quantum gravity effect to the very special relativity, but none of these give rise to noncommutative translations analogous to those of the de Sitter deformation of the Poincar\'e group: space-time remains flat. Among the 2-parameter family of deformation of $ISIM(2)$, they find that only a 1-parameter $DISIM_b (2)$, the deformation of $SIM(2)$, is physically acceptable \cite{Gibbons:2008re}. The line element invariant under $DISIM_b (2)$ is Lorentz violating and of Finsler type, $d{{s}^{2}}={{\left( {{\eta }_{\mu \upsilon }}d{{x}^{\mu }}d{{x}^{\upsilon }} \right)}^{1-b}}{{\left( {{n}_{\mu }}d{{x}^{\mu }} \right)}^{2b}}$. The $DISIM_b (2)$ invariant action for point particle and the wave equations for spin $0$, $\frac{1}{2}$ and $1$ are derived in their paper. The equation for spin $0$ field is in general a nonlocal equation, since it involves fractional even irrational derivatives.

In this paper we follow Gibbons-Gomis-Pope's approach on the deformation of $ISIM(2)$ and investigate the deformation of all such kind of subgroups of Poincar\'e group which are the semi-product of three generators and four generators Lorentz subgroups with the spacetime translation group $T(4)$ (semi-product Poincar\'e subgroup) and the five dimensional representations, which are inherited from the five dimensional representation of Poincar\'e group, (the natural representation) of all the semi-product Poincar\'e subgroup as well as their deformed partners. We find that the deformation of semi-product Poincar\'e subgroup may have more than one families that are physically acceptable. There may be more than one inequivalent natural representations for one family of deformation of a specific Poincar\'e subgroup. Usually the deformation of the original Lorentz subgroup part causes the rotational operation an additional accompanied scale factor which is not reasonable for we believe that the departure from Lorentz symmetry should be from boost rather than rotational operation. Anyhow most deformed boost operations do indeed have an additional accompanied scale factors which will play a key role in the search of group action invariant geometry and construction of field theories in the spactime of the invariant geometry.

\section{Deformation of Lie Algebra}

The deformed Lie algebra or Lie group is extensively investigated \cite{Levy-Nahas:1967da, Gerstenhaber:1964os}. Let's give here a short review on the deformation of Lie algebra according to Gibbons-Gomis-Pope. For a Lie algebra with commutation relations,
\beq \label{eq:comrel}
\left[ {{T}_{i}},{{T}_{j}} \right]=C_{ij}^{k}{{T}_{k}},
\eeq
we can suppose that the structure constants of deformed Lie algebra is of the form
\beq \label{eq:defexpstrcst}
\hat{C}_{ij}^{k}=C_{ij}^{k}+tA_{ij}^{k}+{{t}^{2}}B_{ij}^{k}+... .
\eeq
Here $t$ represents the deformation parameter. The constrain on deformed structure constants from Jacobi identity
\beq \label{eq:jacide}
\left[ \left[ {{T}_{i}},{{T}_{j}} \right],{{T}_{k}} \right]+
\left[ \left[ {{T}_{j}},{{T}_{k}} \right],{{T}_{i}} \right]+
\left[ \left[ {{T}_{k}},{{T}_{i}} \right],{{T}_{j}} \right]=0
\eeq
has the form
\beq \label{eq:jacidecstr}
\hat{C}_{l\left[ k \right.}^{m}\hat{C}_{\left. ij \right]}^{l}=
\hat{C}_{lk}^{m}\hat{C}_{ij}^{l}+\hat{C}_{li}^{m}\hat{C}_{jk}^{l}+
\hat{C}_{lj}^{m}\hat{C}_{ki}^{l}=0.
\eeq
The expansion of deformed structure constant with the power of t yields
\beq \label{eq:perexpjacide}
\begin{array}{c}
  t\left( A_{l\left[ k \right.}^{m}C_{\left. ij \right]}^{l}+
C_{l\left[ k \right.}^{m}A_{\left. ij \right]}^{l} \right)+
  {{t}^{2}}\left( A_{l\left[ k \right.}^{m}A_{\left. ij \right]}^{l}+
B_{l\left[ k \right.}^{m}C_{\left. ij \right]}^{l}+
C_{l\left[ k \right.}^{m}B_{\left. ij \right]}^{l} \right)+...=0.
\end{array}
\eeq
 If there exists a family of deformed Lie algebra parametrized by a continuous variable $t$, there should be a group of constrained equations which arise from every power of $t$ in the above equation, as
 \beq \label{eq:jacidetexplin}
A_{l\left[ k \right.}^{m}C_{\left. ij \right]}^{l}+C_{l\left[ k \right.}^{m}A_{\left. ij \right]}^{l}=0,
\eeq
\beq \label{eq:jacidetexpsec}
 A_{l\left[ k \right.}^{m}A_{\left. ij \right]}^{l}+B_{l\left[ k \right.}^{m}C_{\left. ij \right]}^{l}+C_{l\left[ k \right.}^{m}B_{\left. ij \right]}^{l}=0
\eeq
and etc.

To avoid trivial deformation which arise merely from a change of basis in the original Lie algebra, one demands that there doesn't exist a transformation of basis of Lie algebra $S_{\mu }^{\upsilon }=\delta _{\mu }^{\upsilon }+t\phi _{\mu }^{\upsilon }+...\in GL\left( n, \mathbb{R} \right)$, such that $\hat{C}_{ij}^{k}=S_{c}^{k}C_{ab}^{c}\left( {{S}^{-1}} \right)_{i}^{a}\left( {{S}^{-1}} \right)_{j}^{b}$ and hence
\beq \label{eq:triality}
A_{ij}^{k}=\phi_{l}^{k}C_{ij}^{l}-C_{lj}^{k}\phi _{i}^{l}-C_{il}^{k}\phi _{j}^{l}.
\eeq

Define ${{\lambda }^{\mu }}$ as the basis vector of the original Lie algebra (the left invariant 1-form), then  $d{{\lambda }^{i}}=-\frac{1}{2}C_{ab}^{i}{{\lambda }^{a}}\wedge {{\lambda }^{b}}$[1] [7]. We can define the vector valued one form field ${{\Phi }^{a}}=\phi _{b}^{a}{{\lambda }^{b}}$ and 2-form field ${{A}^{a}}=\frac{1}{2}A_{ij}^{a}{{\lambda }^{i}}\wedge {{\lambda }^{j}}$ and ${{B}^{a}}=\frac{1}{2}B_{ij}^{a}{{\lambda }^{i}}\wedge {{\lambda }^{j}}$ as well as a matrix valued 1-form field $C_{a}^{b}={{\lambda }^{c}}C_{ca}^{b}$. So we have the covariant exterior differential operator of the present Lie algebra $D=d+C\wedge $, the formula \eqref{eq:jacidetexplin} can be rewritten as
\beq \label{eq:extdiffexp}
D{{A}^{a}}=0,{{A}^{a}}\ne -D{{\Phi }^{a}}.
\eeq
The Jacobi Identity requires ${{D}^{2}}=0$, then
\beq \label{eq:secordcstr}
D{{B}^{a}}+{{\left( A\bullet A \right)}^{a}}=0,
\eeq
where ${{\left( A\bullet A \right)}^{a}}=\frac{1}{2}A_{b\left[ c \right.}^{a}A_{\left. de \right]}^{b}{{\lambda }^{c}}\wedge {{\lambda }^{d}}\wedge {{\lambda }^{e}}$. The equation is solvale requires $D{{\left( A\bullet A \right)}^{a}}=0$.

If we set $A\bullet A=0$, we find that the second order term of deformation will also satisfy \eqref{eq:extdiffexp}. Then the acceptable form of ${{B}^{\mu }}$ is the same as one of ${{A}^{\mu }}$. It is enough to consider the first order deformed term only.

\section{The proper subgroups of Lorentz Group}

The Lorentz Lie algebra has the following Lie sub-algebras up to isomorphism.
\begin{itemize}
  \item Lie subalgebra with a single generator
\item two Lie subalgebras with two generators:
$\mathbf{span}\left\{ {{r}_{x}},{{b}_{x}} \right\}$ and $\mathbf{span}\left\{ {{r}_{x}}+{{b}_{y}},{{b}_{z}} \right\}$.
The corresponding commutation relations are
\begin{itemize}
  \item $\mathbf{span}\left\{ {{r}_{x}},{{b}_{x}} \right\}$: $\left[ {{r}_{x}},{{b}_{x}} \right]=0$,
  \item $\mathbf{span}\left\{ {{r}_{x}}+{{b}_{y}},{{b}_{z}} \right\}$: $\left[ {{b}_{x}}+{{r}_{y}},{{b}_{z}} \right]={{b}_{x}}+{{r}_{y}}$.
\end{itemize}
\item four Lie subalgebras with three generators: $\mathbf{span}\left\{ {{r}_{x}},{{r}_{y}},{{r}_{z}} \right\}$, $\mathbf{span}\left\{ {{b}_{x}},{{b}_{y}},{{r}_{z}} \right\}$,$\mathbf{span}\left\{ {{t}_{1}},{{t}_{2}},{{r}_{z}} \right\}$ and $\mathbf{span}\left\{ {{t}_{1}},{{t}_{2}},{{b}_{z}} \right\}$, where ${{t}_{1}}={{b}_{x}}+{{r}_{y}}$ and ${{t}_{2}}={{b}_{y}}-{{r}_{x}}$. The corresponding commutation relations are
\begin{itemize}
  \item $\mathbf{span}\left\{ {{r}_{x}},{{r}_{y}},{{r}_{z}} \right\}$ (the $so(3)$): $\left[ {{r}_{x}},{{r}_{y}} \right]={{r}_{z}},\left[ {{r}_{y}},{{r}_{z}} \right]={{r}_{x}},\left[ {{r}_{z}},{{r}_{x}} \right]={{r}_{y}}$,
  \item $\mathbf{span}\left\{ {{b}_{x}},{{b}_{y}},{{r}_{z}} \right\}$ ( the Lorentz algebra in 2+1 dimension): $\left[ {{b}_{x}},{{b}_{y}} \right]=-{{r}_{z}},\left[ {{b}_{y}},{{r}_{z}} \right]={{b}_{x}},\left[ {{r}_{z}},{{b}_{x}} \right]={{b}_{y}}$.
  \item $\mathbf{span}\left\{ {{t}_{1}},{{t}_{2}},{{r}_{z}} \right\}$ ( the 2 dimensional Eudlidean algebra $e(2)$): $\left[ {{t}_{1}},{{t}_{2}} \right]=0,\left[ {{r}_{z}},{{t}_{1}} \right]={{t}_{2}},\left[ {{r}_{z}},{{t}_{2}} \right]=-{{t}_{1}}$.
  \item $\mathbf{span}\left\{ {{t}_{1}},{{t}_{2}},{{b}_{z}} \right\}$(2-dimensional orientation preserving transformations group $HOM(2)$): $\left[ {{t}_{1}},{{t}_{2}} \right]=0,\left[ {{b}_{z}},{{t}_{1}} \right]=-{{t}_{1}},\left[ {{b}_{z}},{{t}_{2}} \right]=-{{t}_{2}}$.
\end{itemize}
\item one Lie subalgebras with four generators: $\mathbf{span}\left\{ {{t}_{1}},{{t}_{2}},{{r}_{z}},{{b}_{z}} \right\}$ (the 2 dimensional similitude
group $SIM(2)$)
    with commutation relations
    $\left[ {{t}_{1}},{{t}_{2}} \right]=\left[ {{r}_{z}},{{b}_{z}} \right]=0$,
 $\left[ {{r}_{z}},{{t}_{1}} \right]={{t}_{2}},\left[ {{r}_{z}},{{t}_{2}} \right]=-{{t}_{1}}$ and
$\left[ {{b}_{z}},{{t}_{1}} \right]=-{{t}_{1}},\left[ {{b}_{z}},{{t}_{2}} \right]=-{{t}_{2}}$

\end{itemize}

The Lie subalgebra $\mathbf{span}\left\{ {{r}_{x}},{{b}_{x}} \right\}$ is isomorphic to $\mathbf{t} ( 2 ) = \mathbf{span}\left\{ {{t}_{1}},{{t}_{2}} \right\}$, and they are both isomorphic to the 2 dimensional translation group $T(2)$.

We will call the subgroup of Lorentz or Poincar\'e group as Lorentz or Poincar\'e subgroup for brevity.

\section{The Deformation Group of the semi-product subgroups of Poincar\'e Group}

Poincar\'e group is the semi-direct product of Lorentz group with the translation group. Lorantz group is the normal subgroup of the Poincar\'e group which is generated by six generators, three rotation generators ${{r}_{x}},\ {{r}_{y}},\ {{r}_{z}}$ and three boost generators ${{b}_{x}},\ {{b}_{y}},\ {{b}_{z}}$. The semi-direct product of subgroup of Lorentz group with translation group is also the subgroup of Poincar\'e group, which make up one type of Poincar\'e subgroups. We will concentrate our attention on this type of subgroups and it is this type of Poincar\'e subgroup that Cohen and Glashow employ in their very special relativity proposal. The deformation groups of this type of subgroups can also be divided into two kinds. One kind consists of the semi-direct product of the deformation of Lorentz subgroup $SL$ with $T(4)$ , which can be regarded as the locally deformed group, while the deformation group of the other kind does not possess the semi-direct product structure, which can be regarded as the globally deformed group. Among the globally deformed groups, the Lorentz subgroup does not deform in the first class but it will deform in the second class. We will concentrate on the first class of globally deformed groups, in which the deformation part comes from the intercrossing between Lorentz subgroup and the translational group and the translational group itself. The deformed group thus obtained does not have the semi-direct product structure of the Lorentz subgroup with the deformed translation group. Similar to the decomposition of Poincar\'e group into the Lorentz group, the local symmetry group, and the translational group which connect the local properties within a neighborhood, the deformed Poincar\'e subgroups can also be decomposed into two parts, one describe the local properties of the spacetime and the other part reflect the global properties of the spacetime in some extent. We mainly concentrate our attention on that kind of deformed Poincar\'e subgroups in which the Lorentz subgroup part is not deformed so that the local property of spacetime is the same as described in VSR.

From \eqref{eq:secordcstr}, we obtained a constrain condition
\beq \label{eq:nonlincstr}
D{{\left( A\bullet A \right)}^{a}}=0.
\eeq
 The simple solution
\beq \label{eq:solnsecordcstr}
A\bullet A=0
\eeq
  is a solution that satisfies all the constrain condition at all nonlinear orders. Then the  constrain condition from Jacobi Identity \eqref{eq:secordcstr} can be written as
\beq \label{eq:bcstr}
D{{B}^{a}}=0,
\eeq
i.e. the second order deformation of structure constants $B$ satisfies the same equation as $A$. Therefore we can get the higher order of deformation of structure constants in this way. Duo to the simplest solution of the constrain condition \eqref{eq:secordcstr} the deformation of the same group can have several different forms, e.g. the deformation group of $IHOM$, the semidirect product of $HOM$ and $T(4)$, has two different families. Of course the Poincar\'e group itself and $ISIM$ group have only one family of deformation.

\subsection{The Perturbative Solution of the Representation of the Deformed Generators }

  The natural representation of the deformed generators can be viewed as some kind of perturbation of the representation of original group which inherit from the Poincar\'e group's 5 dimensional natural matrix representation for the deformed group can be viewed as the perturbation of the original group. The generators of deformed group can be written as $\left\{ T{{'}_{i}}={{T}_{i}}+\tau {{G}_{i}} \right\}$ and the corresponding structure constants as ${C'}_{ij}^{k}=C_{ij}^{k}+tA_{ij}^{k}$, where $\left\{ {{T}_{i}} \right\}$ and $C_{ij}^{k}$ are the generators and structure constants of the original group, hence
  \beq \label{eq:strcst}
  C_{ij}^{k}{{T}_{k}}=\left[ {{T}_{i}},{{T}_{j}} \right]
  \eeq
  and
  \beq \label{eq:newstrcst}
{C'}_{ij}^{k}T{{'}_{k}}=\left[ T{{'}_{i}},T{{'}_{j}} \right],
  \eeq
 i.e.
  \beq \label{eq: eqrepmatele}
  \begin{array}{c}
    {\tau ^2}\left[ {{G_i},{G_j}} \right] + \tau \left( {\left[ {{G_i},{T_j}} \right] + \left[ {{T_i},{G_j}} \right] - C_{ij}^k{G_k} - tA_{ij}^k{G_k}} \right)
   - tA_{ij}^k{T_k} = 0,
  \end{array}
  \eeq
  where the generators $T$s and $G$s are all $5 \times 5$ matrices and the matrix elements of the unknown $G$s are functions of the deformation parameter $t$. Moreover all of $G$s are zero matrices when $t=0$. We have now $N \times 5 \times 5=25N$ unknown variables for a Lie algebra with $N$ generators, e.g. there are 250 unknown variables for Poincar\'e group, 200 for $ISIM$ group and 175 for $IHOM$ group respectively.

We can solve \eqref{eq: eqrepmatele} perturbatively. The dominant part of perturbation parameter $\tau$ for generators and $t$ for structure constants should be in the same order. In general, we can assume that $tA_{ij}^k = \tau \bar A_{ij}^k $. \eqref{eq: eqrepmatele} becomes
\begin{equation}\label{eq:pertexpmatrep}
\left\{ \begin{array}{l}
\left[ {{G_i},{G_j}} \right] - \bar A_{ij}^k{G_k}  = 0\\
\left[ {{G_i},{T_j}} \right] + \left[ {{T_i},{G_j}} \right] - C_{ij}^k{G_k} - \bar A_{ij}^k{T_k} = 0
\end{array} \right.\
\end{equation}

The simplest case is $t_1 A_{ij}^k=\bar A_{ij}^k$ and $t =  {t_1}\tau $.
Rewrite $t_1$ as $t$, finally we have
\begin{equation}\label{eq:texpmatrep}
\left\{ \begin{array}{l}
\left[ {{G_i},{G_j}} \right] - tA_{ij}^k{G_k}  = 0\\
\left[ {{G_i},{T_j}} \right] + \left[ {{T_i},{G_j}} \right] - C_{ij}^k{G_k} - tA_{ij}^k{T_k} = 0.
\end{array} \right.
\end{equation}

There may be more than one set of solutions due to the quadratic equations. We find that there may be more than one inequivalent natural representations for the deformation of a specific Lie algebra, which corresponding to different spacetime geometry.

\subsection{The deformation of Poincar\'e group}

The commutation relations for Poincar\'e group are
\beq \label{eq: commrelpoincare}
\begin{array}{c}
  \left[ {{r_i},{r_j}} \right] = \sum\limits_{k = 1}^3 {{\varepsilon _{ijk}}{r_k}}, \left[ {{b_i},{b_j}} \right] =  - \sum\limits_{k = 1}^3 {{\varepsilon _{ijk}}{r_k}}, \left[ {{b_i},{r_j}} \right] = \sum\limits_{k = 1}^3 {{\varepsilon _{ijk}}{b_k}}, \\
  \left[ {{p_i},{p_j}} \right] = 0,\left[ {{r_i},{p_t}} \right] = 0,
  \left[ {{r_i},{p_j}} \right] = \sum\limits_{k = 1}^3 {{\varepsilon _{ijk}}{p_k}}, \left[ {{b_i},{p_t}} \right] = {p_i}, \left[ {{b_i},{p_j}} \right] = {\delta _{ij}}{p_t}.
\end{array}
\eeq

The first order Jacobi constrain equation,
\[A_{l\left[ k \right.}^mC_{\left. {ij} \right]}^l + C_{l\left[ k \right.}^mA_{\left. {ij} \right]}^l = 0, \]
the simplest solution $A \bullet A = 0$ as the second order constrain, and the non-triviality condition,
\[A_{ij}^k \ne \phi _l^kC_{ij}^l - C_{lj}^k\phi _i^l - C_{il}^k\phi _j^l, \]
reduce most of the possible $10 \times \frac{10\times 9}{2}=450$ deformation parameters $A^i_{jk}$ to zero and it can be verified that the deformation group of Poincar\'e group is unique and possesses the commutation structure,
\beq \label{eq: defcommrelpoincare}
\begin{array}{c}
  \left[ {{r_i},{r_j}} \right] = \sum\limits_{k = 1}^3 {{\varepsilon _{ijk}}{r_k}}, \left[ {{b_i},{b_j}} \right] =  - \sum\limits_{k = 1}^3 {{\varepsilon _{ijk}}{r_k}}, \left[ {{b_i},{r_j}} \right] = \sum\limits_{k = 1}^3 {{\varepsilon _{ijk}}{b_k}}, \left[ {{r_i},{p_t}} \right] = 0, \\

 \left[ {{r_i},{p_j}} \right] = \sum\limits_{k = 1}^3 {{\varepsilon _{ijk}}{p_k}}, \left[ {{b_i},{p_t}} \right] = {p_i},
  \left[ {{b_i},{p_j}} \right] = {\delta _{ij}}{p_t}, \left[ {{p_t},{p_i}} \right] = t{b_i}, \left[ {{p_i},{p_j}} \right] =  - t\sum\limits_{k = 1}^3 {{\varepsilon _{ijk}}{r_k}},
\end{array}
\eeq
which is known as Lie algebra of de Sitter group.

The natural representation of the generators is also unique, which has the form,
\beq \label{eq:repmatdesitter}
\begin{array}{c}
  {p_t} = \left( {\begin{array}{*{20}{c}}
{}&{}&{}&{}&1\\
{}&{}&{}&{}&{}\\
{}&{}&{}&{}&{}\\
{}&{}&{}&{}&{}\\
{ - t}&{}&{}&{}&{}
\end{array}} \right),{p_x} = \left( {\begin{array}{*{20}{c}}
{}&{}&{}&{}&{}\\
{}&{}&{}&{}&1\\
{}&{}&{}&{}&{}\\
{}&{}&{}&{}&{}\\
{}&t&{}&{}&{}
\end{array}} \right),
  {p_y} = \left( {\begin{array}{*{20}{c}}
{}&{}&{}&{}&{}\\
{}&{}&{}&{}&{}\\
{}&{}&{}&{}&1\\
{}&{}&{}&{}&{}\\
{}&{}&t&{}&{}
\end{array}} \right),{p_z} = \left( {\begin{array}{*{20}{c}}
{}&{}&{}&{}&{}\\
{}&{}&{}&{}&{}\\
{}&{}&{}&{}&{}\\
{}&{}&{}&{}&1\\
{}&{}&{}&t&{}
\end{array}} \right),
\end{array}
\eeq
where we only denote the non-zero matrix elements of the deformed generators, i.e. the representation matrix of the other six generators, the generators of the Lorentz group, remain unchanged.

\subsection{The deformation of $ISIM$}

The algebraic structure of $ISIM$, the semi-product of $SIM$ with $T(4)$, is
\beq \label{eq: commrelisim}
\begin{array}{c}
  \left[ {{t_1},{r_z}} \right] =  - {t_2},  \left[ {{t_1},{b_z}} \right] = {t_1},  \left[ {{t_1},{p_t}} \right] = \left[ {{t_1},{p_z}} \right] = {p_x},
  \left[ {{t_2},{r_z}} \right] = {t_1}, \left[ {{t_2},{b_z}} \right] = {t_2}, \left[ {{t_2},{p_t}} \right] = \left[ {{t_2},{p_z}} \right] = {p_y}, \\
  \left[ {{t_1},{p_x}} \right] = {p_t} - {p_z},\left[ {{t_2},{p_y}} \right] = {p_t} - {p_z},\left[ {{r_z},{p_x}} \right] = {p_y},
  \left[ {{r_z},{p_y}} \right] =  - {p_x} , \left[ {{b_z},{p_t}} \right] = {p_z} , \left[ {{b_z},{p_z}} \right] = {p_t}.
\end{array}
\eeq
The Jacobi constrain reduces the $8 \times \frac{8\times 7}{2}=224$ deformation parameters of the deformed group $DISIM$ to 57. The simplest solution $A \bullet A = 0$ then reduce further to 6 ones,
\beq \label{eq: defparadisim}
A_{1b}^1,A_{1x}^t,A_{1x}^z,A_{rt}^t,A_{bt}^t,A_{bt}^z,
\eeq
where $r, b,t,x,z$ represent ${r_z},{b_z},{p_t},{p_x},{p_z}$ respectively. The commutation relation for $DISIM$ is
\beq \label{eq: commreldisim}
\begin{array}{c}
  \left[ {{t_1},{r_z}} \right] =  - {t_2}, \left[ {{t_1},{b_z}} \right] = \left( {1 + A_{1b}^1} \right){t_1}, \left[ {{t_2},{r_z}} \right] = {t_1},
  \left[ {{t_1},{p_t}} \right] = {p_x}, \left[ {{t_1},{p_x}} \right] = \left( {1 + A_{1x}^t} \right){p_t} - \left( {1 - A_{1x}^z} \right){p_z}, \\
  \left[ {{t_1},{p_z}} \right] = \left( {1 + A_{1x}^t + A_{1x}^z} \right){p_x}, \left[ {{t_2},{b_z}} \right] = \left( {1 + A_{1b}^1} \right){t_2},
  \left[ {{t_2},{p_t}} \right] = {p_y},\left[ {{t_2},{p_y}} \right] = \left( {1 + A_{1x}^t} \right){p_t} - \left( {1 - A_{1x}^z} \right){p_z}, \\
  \left[ {{t_2},{p_z}} \right] = \left( {1 + A_{1x}^t + A_{1x}^z} \right){p_y}, \left[ {{r_z},{p_t}} \right] = A_{rt}^t{p_t} ,
  \left[ {{r_z},{p_x}} \right] = {p_y} + A_{rt}^t{p_x} , \left[ {{r_z},{p_y}} \right] =  - {p_x} + A_{rt}^t{p_y} , \\
  \left[ {{r_z},{p_z}} \right] = A_{rt}^t{p_z},
  \left[ {{b_z},{p_x}} \right] = \left( {A_{1x}^t + A_{1x}^z + A_{bt}^t + A_{bt}^z - A_{1b}^1} \right){p_x},
  \left[ {{b_z},{p_y}} \right] = \left( {A_{1x}^t + A_{1x}^z + A_{bt}^t + A_{bt}^z - A_{1b}^1} \right){p_y}, \\
  \left[ {{b_z},{p_t}} \right] = {p_z} + A_{bt}^t{p_t} + A_{bt}^z{p_z} , \left[ {{b_z},{p_z}} \right] = {p_t} + \left( {2A_{1b}^1 - A_{bt}^z} \right){p_t}
  + \left( {2A_{1x}^t + 2A_{1x}^z + A_{bt}^t + 2A_{bt}^z - 2A_{1b}^1} \right){p_z}.
\end{array}
\eeq
The non-triviality condition is
\beq \label{eq: ntrvdisim}
{A{_{rt}^t}^2} + {\left( {A_{1x}^t + A_{1x}^z + A_{bt}^t + A_{bt}^z - A_{1b}^1} \right)^2} \ne 0.
\eeq
The simplest solution $A \bullet A = 0$ gives
\beq \label{eq: eqefpardisim}
\left\{ \begin{array}{l}
A_{1x}^z\left( {A_{1x}^t + A_{1x}^z} \right) = 0\\
A_{bt}^z\left( {A_{1x}^t + A_{1x}^z} \right) = 0\\
\left( {A_{1x}^t - 2A_{1b}^1} \right)\left( {A_{1x}^t + A_{1x}^z} \right) = 0
\end{array} \right.
\eeq
The existence of deformation parameter $A_{1b}^1$ reveals that there is deformation inside of the original $sim$ Lie subalgebra. We thus can specify $DISIM$ into two families.

\subsubsection{The deformation group with $SIM$ undeformed}

If $A_{1b}^1=0$, $SIM$ is undeformed in $DISIM$ from \eqref{eq: commreldisim}. The non-triviality condition now reads
\beq \label{eq: nontrivdisim1}
{A{_{rt}^t}^2} + {\left( {A_{1x}^t + A_{1x}^z + A_{bt}^t + A_{bt}^z} \right)^2} \ne 0.
\eeq
The quadratic constrain condition becomes
\beq \label{eq: parcstrdisim1}
\left\{ \begin{array}{l}
A_{1x}^z\left( {A_{1x}^t + A_{1x}^z} \right) = 0\\
A_{bt}^z\left( {A_{1x}^t + A_{1x}^z} \right) = 0\\
A_{1x}^t\left( {A_{1x}^t + A_{1x}^z} \right) = 0
\end{array} \right..
\eeq

From \eqref{eq: parcstrdisim1}, the deformation group with $SIM$ undeformed can be classified into two subfamilies: 1, $A_{1x}^z =  - A_{1x}^t$, and 2, $A_{1x}^z = A_{bt}^z = A_{1x}^t = 0$.

In the first subfamily, $A_{1x}^t$ can be absorbed into the redefinition of the generators,
\beq \label{eq: genreddisim1}
\left\{ \begin{array}{l}
{t_i} \to {\left( {1 + A_{1x}^t} \right)^{ - {1 \mathord{\left/
 {\vphantom {1 2}} \right.
 \kern-\nulldelimiterspace} 2}}}{t_i},\;i = 1,2 \\
{p_\alpha } \to {\left( {1 + A_{1x}^t} \right)^{{1 \mathord{\left/
 {\vphantom {1 2}} \right.
 \kern-\nulldelimiterspace} 2}}}{p_\alpha },\alpha  = t,z
\end{array} \right..
\eeq

There are three deformation parameters left, $A_{rt}^t,A_{bt}^t,A_{bt}^z$, which can be simplified further. In fact, any $A^t_{bt}$ gives the same Lie algebra upto an isomorphism when $A_{bt}^t + A_{bt}^z$ is kept fixed. For example, there are two Lie algebras, $t_1^{\left( i \right)},t_2^{\left( i \right)},r_z^{\left( i \right)},b_z^{\left( i \right)},p_t^{\left( i \right)},p_x^{\left( i \right)},p_y^{\left( i \right)},p_z^{\left( i \right)}$ where $i=1$ corresponds to one set of deformation parameters $A_{rt}^t,A_{bt}^t,A_{bt}^z$ and $i=2$ corresponds to the other set of deformation parameters $A_{rt}^t,B_{bt}^t,B_{bt}^z$ satisfying $A_{bt}^t + A_{bt}^z = B_{bt}^t + B_{bt}^z$. We then can define
\beq \label{eq: exmgenreddisim1}
\left\{ \begin{array}{l}
 p_t^{\left( 2 \right)} = p_t^{\left( 1 \right)} + \frac{1}{2}\left( {A_{bt}^t - B_{bt}^t} \right)\left( {p_t^{\left( 1 \right)} - p_z^{\left( 1 \right)}} \right)\\
p_z^{\left( 2 \right)} = p_z^{\left( 1 \right)} + \frac{1}{2}\left( {A_{bt}^t - B_{bt}^t} \right)\left( {p_t^{\left( 1 \right)} - p_z^{\left( 1 \right)}} \right)
\end{array} \right..
\eeq
such that $p_t^{\left( 2 \right)} - p_z^{\left( 2 \right)} = p_t^{\left( 1 \right)} - p_z^{\left( 1 \right)}$ and
\beq \label{eq: exmgenredcommrel}
\begin{array}{l}
\left[ {{b_z},p_t^{\left( 2 \right)}} \right] = \left[ {{b_z},p_t^{\left( 1 \right)}} \right] + \frac{{A_{bt}^t - B_{bt}^t}}{2}\left[ {{b_z},p_t^{\left( 1 \right)} - p_z^{\left( 1 \right)}} \right]\\
 = p_z^{\left( 1 \right)} + A_{bt}^tp_t^{\left( 1 \right)} + A_{bt}^zp_z^{\left( 1 \right)} + \frac{{A_{bt}^t - B_{bt}^t}}{2}\left( { - 1 + A_{bt}^t + A_{bt}^z} \right)\left( {p_t^{\left( 1 \right)} - p_z^{\left( 1 \right)}} \right)\\
 = p_z^{\left( 1 \right)} - \frac{{A_{bt}^t - B_{bt}^t}}{2}\left( {p_t^{\left( 1 \right)} - p_z^{\left( 1 \right)}} \right) + A_{bt}^tp_t^{\left( 2 \right)} + A_{bt}^zp_z^{\left( 2 \right)}
 = p_z^{\left( 2 \right)} - \left( {A_{bt}^t - B_{bt}^t} \right)\left( {p_t^{\left( 2 \right)} - p_z^{\left( 2 \right)}} \right) + A_{bt}^tp_t^{\left( 2 \right)} + A_{bt}^zp_z^{\left( 2 \right)}\\
 = p_z^{\left( 2 \right)} + B_{bt}^tp_t^{\left( 2 \right)} + \left( {A_{bt}^t + A_{bt}^z - B_{bt}^t} \right)p_z^{\left( 2 \right)}.
\end{array}
\eeq

We therefore only consider two cases in which $A_{bt}^t = 0$ or $A_{bt}^z = 0$.

In the second subfamily, there are two deformation parameters $A^t_{rt}$ and $A^t_{bt}$, and therefore it can be classified into the first subfamily.

There remain two cases to be investigated, $A_{bt}^t = 0$ for the first case and $A_{bt}^z = 0$ for the second case.

Let's consider the first case in which $A_{bt}^t = 0$. Denoting $A_1 =A^t_{rt}$ and $A_2 =A^t_{bt}$, the representation matrices of the deformed generators are
\beq \label{eq: matrepdisim1}
\begin{array}{c}
{r_z} = \left( {\begin{array}{*{20}{c}}
{{A_1}}&{}&{}&{}&{}\\
{}&{{A_1}}&{ - 1}&{}&{}\\
{}&1&{{A_1}}&{}&{}\\
{}&{}&{}&{{A_1}}&{}\\
{}&{}&{}&{}&0
\end{array}} \right),
{b_z} = \left( {\begin{array}{*{20}{c}}
{{A_2}}&{}&{}&1&{}\\
{}&{{A_2}}&{}&{}&{}\\
{}&{}&{{A_2}}&{}&{}\\
1&{}&{}&{{A_2}}&{}\\
{}&{}&{}&{}&0
\end{array}} \right),
\end{array}
\eeq
and the corresponding single parameter group elements are
\beq \label{eq: grpeleDISIM1}
\begin{array}{l}
{R_z}\left( \theta  \right) = \left( {\begin{array}{*{20}{c}}
{{e^{\theta {A_1}}}}&{}&{}&{}&{}\\
{}&{{e^{\theta {A_1}}}\cos \theta }&{ - {e^{\theta {A_1}}}\sin \theta }&{}&{}\\
{}&{{e^{\theta {A_1}}}\sin \theta }&{{e^{\theta {A_1}}}\cos \theta }&{}&{}\\
{}&{}&{}&{{e^{\theta {A_1}}}}&{}\\
{}&{}&{}&{}&1
\end{array}} \right),
{B_z}\left( \theta  \right) = \left( {\begin{array}{*{20}{c}}
{{e^{\theta {A_2}}}\cosh \theta }&{}&{}&{{e^{\theta {A_2}}}\sinh \theta }&{}\\
{}&{{e^{\theta {A_2}}}}&{}&{}&{}\\
{}&{}&{{e^{\theta {A_2}}}}&{}&{}\\
{{e^{\theta {A_2}}}\sinh \theta }&{}&{}&{{e^{\theta {A_2}}}\cosh \theta }&{}\\
{}&{}&{}&{}&1
\end{array}} \right),
\end{array}
\eeq
where the deformed rotation $R_z\left( \theta  \right)$ is not a merely rotation anymore but a rotation followed by a dilatation $e^{\theta {A_1}}$. $R_z\left( 2\pi  \right)=e^{2\pi {A_1}}$ is a pure dilatation when $A_1 \neq 0$. To keep $R_z\left( \theta  \right)$ as a reasonable local rotation operation, one demands $A_1 = 0$. There survive only one deformation parameter $A_2$, denoted by $b$ hereafter, for this case. The representation matrix of the deformed boost operation is now of the form,
\beq \label{eq: boostDISIM1}
{B_z}\left( \theta  \right) = {e^{b\theta }}\left( {\begin{array}{*{20}{c}}
{\cosh \theta }&{}&{}&{\sinh \theta }\\
{}&1&{}&{}\\
{}&{}&1&{}\\
{\sinh \theta }&{}&{}&{\cosh \theta }
\end{array}} \right),
\eeq
an ordinary boost followed by a dilatation.

In the second case, $A_{bt}^z = 0$. Denoting $A_1 =A^t_{rt}$ and $A_2 =A^z_{bt}$, what is different from the first case just investigated is that there may exists a group of matrix representation for the deformed group which is specified by a free parameter $\lambda$:
\beq \label{eq: repgrpcase2disim1}
\begin{array}{l}
{r_z} = \left( {\begin{array}{*{20}{c}}
{{A_1}}&{}&{}&{}&{}\\
{}&{{A_1}}&{ - 1}&{}&{}\\
{}&1&{{A_1}}&{}&{}\\
{}&{}&{}&{{A_1}}&{}\\
{}&{}&{}&{}&0
\end{array}} \right),
{b_z} = \left( {\begin{array}{*{20}{c}}
{2\lambda }&{}&{}&{1 - {A_2} + 2\lambda }&{}\\
{}&{{A_2}}&{}&{}&{}\\
{}&{}&{{A_2}}&{}&{}\\
{1 + {A_2} - 2\lambda }&{}&{}&{2\left( {{A_2} - \lambda } \right)}&{}\\
{}&{}&{}&{}&0
\end{array}} \right),\\
{p_t} = \left( {\begin{array}{*{20}{c}}
0&{}&{}&{}&{1 + \lambda }\\
{}&0&{}&{}&{}\\
{}&{}&0&{}&{}\\
{}&{}&{}&0&{ - \lambda }\\
{}&{}&{}&{}&0
\end{array}} \right),
{p_z} = \left( {\begin{array}{*{20}{c}}
0&{}&{}&{}&\lambda \\
{}&0&{}&{}&{}\\
{}&{}&0&{}&{}\\
{}&{}&{}&0&{1 - \lambda }\\
{}&{}&{}&{}&0
\end{array}} \right).
\end{array}
\eeq

Similar to the first case, one can arrive at a reasonable local rotation operation by forcing the rotation generator undeformed. The free parameter $\lambda$ actually represents the choice of coordinate system. It means that the representation matrices which different $\lambda$ corresponds to can be transformed from one to another by a coordinate transformation, e.g. the matrix representation of $\lambda =\lambda_1$ can be transformed to ones of $\lambda =\lambda_2$ by the following coordinate transformation matrix,
\beq \label{eq: trsfmat1}
T = \left( {\begin{array}{*{20}{c}}
{1 - {\lambda _1} + {\lambda _2}}&{}&{}&{{\lambda _2} - {\lambda _2}}&{}\\
{}&1&{}&{}&{}\\
{}&{}&1&{}&{}\\
{{\lambda _1} - {\lambda _2}}&{}&{}&{1 + {\lambda _1} - {\lambda _2}}&{}\\
{}&{}&{}&{}&1
\end{array}} \right).
\eeq
What we need is therefore to choose an appropriate $\lambda$, e.g. $\lambda=\frac{A_2}{2}$, and the representation matrices for generators are
\beq \label{eq: simrepmatdisim1}
\begin{array}{l}
{b_z} = \left( {\begin{array}{*{20}{c}}
{{A_2}}&{}&{}&1&{}\\
{}&{{A_2}}&{}&{}&{}\\
{}&{}&{{A_2}}&{}&{}\\
1&{}&{}&{{A_2}}&{}\\
{}&{}&{}&{}&0
\end{array}} \right),
{p_t} = \left( {\begin{array}{*{20}{c}}
0&{}&{}&{}&{1 + \frac{{{A_2}}}{2}}\\
{}&0&{}&{}&{}\\
{}&{}&0&{}&{}\\
{}&{}&{}&0&{ - \frac{{{A_2}}}{2}}\\
{}&{}&{}&{}&0
\end{array}} \right),
{p_t} = \left( {\begin{array}{*{20}{c}}
0&{}&{}&{}&{\frac{{{A_2}}}{2}}\\
{}&0&{}&{}&{}\\
{}&{}&0&{}&{}\\
{}&{}&{}&0&{1 - \frac{{{A_2}}}{2}}\\
{}&{}&{}&{}&0
\end{array}} \right).
\end{array}
\eeq
The corresponding single parameter group elements are
\beq \label{eq: simgrpeledisim1}
\begin{array}{l}
{B_z}\left( \theta  \right) = {e^{b\theta }}\left( {\begin{array}{*{20}{c}}
{\cosh \theta }&{}&{}&{\sinh \theta }\\
{}&1&{}&{}\\
{}&{}&1&{}\\
{\sinh \theta }&{}&{}&{\cosh \theta }
\end{array}} \right),
{P_t}\left( \lambda  \right) = \left( {\begin{array}{*{20}{c}}
{\lambda  + \frac{{{A_2}}}{2}\lambda }\\
{}\\
{}\\
{ - \frac{{{A_2}}}{2}\lambda }
\end{array}} \right),
{P_z}\left( \lambda  \right) = \left( {\begin{array}{*{20}{c}}
{\frac{{{A_2}}}{2}\lambda }\\
{}\\
{}\\
{\lambda  - \frac{{{A_2}}}{2}\lambda }
\end{array}} \right).
\end{array}
\eeq
Note that there are many different matrix representations as a matter of fact. However, the $5\times 5$ representation matrices of the deformed group elements have their origin from the $5\times 5$ representation of Poincar\'e group which has a special geometric explanation. The $5\times 5$ representation of the deformed group should have the same geometric explanation, i.e. the upper left $4\times 4$ part of the representation matrix represents rotation and boost, the upper right $1\times 4$ part represents translation and the lower $5\times 1$ part should keep to be zero. The following matrix representation of the first subclass is excluded with this restriction,
\beq \label{eq: ignrepmatdisim1}
\begin{array}{c}
{r_z} = \left( {\begin{array}{*{20}{c}}
0&{}&{}&{}&{}\\
{}&0&{ - 1}&{}&{}\\
{}&1&0&{}&{}\\
{}&{}&{}&0&{}\\
{}&{}&{}&{}&{ - {A_1}}
\end{array}} \right),
{b_z} = \left( {\begin{array}{*{20}{c}}
0&{}&{}&1&{}\\
{}&0&{}&{}&{}\\
{}&{}&0&{}&{}\\
1&{}&{}&0&{}\\
{}&{}&{}&{}&{ - {A_2}}
\end{array}} \right),
\end{array}
\eeq

which do not have an apparent geometric explanation. We will ignore such kind of representation hereafter.
\subsubsection{The deformation group with $SIM$ deformed}
In the last section we have investigated the deformation group in which the $SIM$ part remains un-deformed and the corresponding natural representation. We are going to investigate the deformation of $ISIM$ in which the $SIM$ itself also deforms and the corresponding natural representation in this section.
Like the case where $SIM$ is undeformed, we can specify two subfamilies, 1. $A_{1x}^z =  - A_{1x}^t$ and 2. $A_{1x}^z = A_{bt}^z = 0$ and $A_{1x}^t = 2A_{1b}^1$.

The first subfamily is denoted by $xdisim1$, in which there are 4 deform parameters, $A_{1b}^1,A_{rt}^t,A_{bt}^t,A_{bt}^z$, where the deformed Lie algebra with arbitrary value of $A^t_{bt}$ is the same one up to  an isomorphism only if $A_{bt}^t + A_{bt}^z$ is kept fixed. There are three independent deform parameters, $A_{1b}^1,A_{rt}^t$ and $A_{bt}^t + A_{bt}^z$ actually.

We can also specify two cases further as in last section. In the first case , the independent deform parameters are $A_{1b}^1,A_{rt}^t$ and $A_{bt}^t$, and the commutation relations are
\beq \label{eq: commrelxisim1}
\begin{array}{l}
\left[ {{t_1},{b_z}} \right] = \left( {1 + A_{1b}^1} \right){t_1}, \left[ {{t_2},{b_z}} \right] = \left( {1 + A_{1b}^1} \right){t_2},
\left[ {{r_z},{p_t}} \right] = A_{rt}^t{p_t},
\left[ {{r_z},{p_x}} \right] = {p_y} + A_{rt}^t{p_x},
\left[ {{r_z},{p_y}} \right] =  - {p_x} + A_{rt}^t{p_y},
 \left[ {{r_z},{p_z}} \right] = A_{rt}^t{p_z},\\
\left[ {{b_z},{p_t}} \right] = {p_z} + A_{bt}^t{p_t}, \left[ {{b_z},{p_x}} \right] = \left( {A_{bt}^t - A_{1b}^1} \right){p_x},
\left[ {{b_z},{p_y}} \right] = \left( {A_{bt}^t - A_{1b}^1} \right){p_y},
\left[ {{b_z},{p_z}} \right] = {p_t} + 2A_{1b}^1{p_t} + \left( {A_{bt}^t - 2A_{1b}^1} \right){p_z}.
\end{array}
\eeq
The natural matrix representation are
\beq \label{eq: marrepxisim1}
\begin{array}{l}
{b_z} = \left( {\begin{array}{*{20}{c}}
{\alpha  - 2{A_1} + {A_3}}&{}&{}&{1 + \alpha }\\
{}&{{A_3} - {A_1}}&{}&{}\\
{}&{}&{{A_3} - {A_1}}&{}\\
{1 - \alpha  + 2{A_1}}&{}&{}&{{A_3} - \alpha }
\end{array}} \right),
{r_z} = \left( {\begin{array}{*{20}{c}}
{{A_2}}&{}&{}&{}\\
{}&{{A_2}}&{ - 1}&{}\\
{}&1&{{A_2}}&{}\\
{}&{}&{}&{{A_2}}
\end{array}} \right), \\
{p_t} = \left( {\begin{array}{*{20}{c}}
{1 + \frac{\alpha }{2}}\\
{}\\
{}\\
{{A_1} - \frac{\alpha }{2}}
\end{array}} \right),
{p_x} = \left( {\begin{array}{*{20}{c}}
{}\\
{1 + {A_1}}\\
{}\\
{}
\end{array}} \right),
{p_y} = \left( {\begin{array}{*{20}{c}}
{}\\
{}\\
{1 + {A_1}}\\
{}
\end{array}} \right),
{p_z} = \left( {\begin{array}{*{20}{c}}
{\frac{\alpha }{2} - {A_1}}\\
{}\\
{}\\
{1 + 2{A_1} - \frac{\alpha }{2}}
\end{array}} \right),
\end{array}
\eeq
where $\alpha$ is a free parameter such that the matrix representations of different value of which can be transformed from one to another. The transformation matrix
\beq \label{eq: trfmat2}
T = \left( {\begin{array}{*{20}{c}}
{1 + \frac{{{\alpha _2} - {\alpha _1}}}{{2 + 2{A_1}}}}&{}&{}&{\frac{{{\alpha _2} - {\alpha _1}}}{{2 + 2{A_1}}}}&{}\\
{}&1&{}&{}&{}\\
{}&{}&1&{}&{}\\
{ - \frac{{{\alpha _2} - {\alpha _1}}}{{2 + 2{A_1}}}}&{}&{}&{1 - \frac{{{\alpha _2} - {\alpha _1}}}{{2 + 2{A_1}}}}&{}\\
{}&{}&{}&{}&1
\end{array}} \right)
\eeq
can transform the matrix representation of $\alpha =\alpha_1$ to one of $\alpha =\alpha_2$. So that we can give $\alpha$ a suitable value ,e.g. $\alpha =A_1$ and therefore
\beq \label{eq: smprepxisim1}
\begin{array}{l}
{b_z} = \left( {\begin{array}{*{20}{c}}
{{A_3} - {A_1}}&{}&{}&{1 + {A_1}}\\
{}&{{A_3} - {A_1}}&{}&{}\\
{}&{}&{{A_3} - {A_1}}&{}\\
{1 + {A_1}}&{}&{}&{{A_3} - {A_1}}
\end{array}} \right),
{r_z} = \left( {\begin{array}{*{20}{c}}
{{A_2}}&{}&{}&{}\\
{}&{{A_2}}&{ - 1}&{}\\
{}&1&{{A_2}}&{}\\
{}&{}&{}&{{A_2}}
\end{array}} \right), \\
{p_t} = \left( {\begin{array}{*{20}{c}}
{1 + \frac{{{A_1}}}{2}}\\
{}\\
{}\\
{\frac{{{A_1}}}{2}}
\end{array}} \right),
{p_x} = \left( {\begin{array}{*{20}{c}}
{}\\
{1 + {A_1}}\\
{}\\
{}
\end{array}} \right),
{p_y} = \left( {\begin{array}{*{20}{c}}
{}\\
{}\\
{1 + {A_1}}\\
{}
\end{array}} \right), {p_z} = \left( {\begin{array}{*{20}{c}}
{ - \frac{{{A_1}}}{2}}\\
{}\\
{}\\
{1 + \frac{{3{A_1}}}{2}}
\end{array}} \right).
\end{array}
\eeq

The corresponding single parameter group elements are
\beq \label{eq: grpelexisim1}
\begin{array}{l}
{B_z}\left( \theta  \right) = {e^{\theta \left( {{A_3} - {A_1}} \right)}}\left( {\begin{array}{*{20}{c}}
{\cosh \left( {1 + {A_1}} \right)\theta }&{}&{}&{\sinh \left( {1 + {A_1}} \right)\theta }\\
{}&1&{}&{}\\
{}&{}&1&{}\\
{\sinh \left( {1 + {A_1}} \right)\theta }&{}&{}&{\cosh \left( {1 + {A_1}} \right)\theta }
\end{array}} \right),
{P_t}\left( \lambda  \right) = \left( {\begin{array}{*{20}{c}}
{\lambda  + \frac{{{A_1}}}{2}\lambda }\\
{}\\
{}\\
{\frac{{{A_1}}}{2}\lambda }
\end{array}} \right),
{P_z}\left( \lambda  \right) = \left( {\begin{array}{*{20}{c}}
{ - \frac{{{A_1}}}{2}\lambda }\\
{}\\
{}\\
{\lambda  + \frac{{3{A_1}}}{2}\lambda }
\end{array}} \right).
\end{array}
\eeq

In the second case of $xdisim1$, the deform parameters are $A_{1b}^1,A_{rt}^t,A_{bt}^z$ and the commutation relations are
\beq \label{eq: commrel2xisim1}
\begin{array}{l}
\left[ {{t_1},{b_z}} \right] = \left( {1 + A_{1b}^1} \right){t_1},  \left[ {{t_2},{b_z}} \right] = \left( {1 + A_{1b}^1} \right){t_2},
\left[ {{r_z},{p_t}} \right] = A_{rt}^t{p_t},  \left[ {{r_z},{p_x}} \right] = {p_y} + A_{rt}^t{p_x},\\
\left[ {{r_z},{p_y}} \right] =  - {p_x} + A_{rt}^t{p_y},
 \left[ {{r_z},{p_z}} \right] = A_{rt}^t{p_z},
\left[ {{b_z},{p_t}} \right] = {p_z} + A_{bt}^z{p_z}, \left[ {{b_z},{p_x}} \right] = \left( {A_{bt}^z - A_{1b}^1} \right){p_x},\\
\left[ {{b_z},{p_y}} \right] = \left( {A_{bt}^z - A_{1b}^1} \right){p_y},
\left[ {{b_z},{p_z}} \right] = {p_t} + \left( {2A_{1b}^1 - A_{bt}^z} \right){p_t} + 2\left( {A_{bt}^z - A_{1b}^1} \right){p_z}.
\end{array}
\eeq
There are many equivalent representations and we can choose a simple one as in the first case,
\beq \label{eq: matrep2xisim1}
\begin{array}{l}
{r_z} = \left( {\begin{array}{*{20}{c}}
{{A_2}}&{}&{}&{}&{}\\
{}&{{A_2}}&{ - 1}&{}&{}\\
{}&1&{{A_2}}&{}&{}\\
{}&{}&{}&{{A_2}}&{}\\
{}&{}&{}&{}&0
\end{array}} \right),
{b_z} =\left( {\begin{array}{*{20}{c}}
0&{}&{}&{1 - {A_3} + 2{A_1}}&{}\\
{}&{{A_3} - {A_1}}&{}&{}&{}\\
{}&{}&{{A_3} - {A_1}}&{}&{}\\
{1 + {A_3}}&{}&{}&{2\left( {{A_3} - {A_1}} \right)}&{}\\
{}&{}&{}&{}&0
\end{array}} \right),
\end{array}
\eeq
where $A_i$ represent $A_{1b}^1,A_{rt}^t,A_{bt}^z$. For the same reason as in the last section, the reasonable request that a local rotation operation should not have an additional dilatation transformation constrains $A_2=0$. Hence the deformed group element is
\beq \label{eq: boostdef2xisim1}
\begin{array}{c}
{B_z}\left( \theta  \right) = {e^{\theta \left( {{A_3} - {A_1}} \right)}}\left( {\begin{array}{*{20}{c}}\\
{\cosh \omega  + \frac{{{A_1} - {A_3}}}{{1 + {A_1}}}\sinh \omega }&{}&{}&{\frac{{1 + 2{A_1} - {A_3}}}{{1 + {A_1}}}\sinh \omega }\\
{}&1&{}&{}\\
{}&{}&1&{}\\
{\frac{{1 + {A_3}}}{{1 + {A_1}}}\sinh \omega }&{}&{}&{\cosh \omega  - \frac{{{A_1} - {A_3}}}{{1 + {A_1}}}\sinh \omega }
\end{array}} \right),
\end{array}
\eeq
where $\omega  = \left( {1 + {A_1}} \right)\theta$. Note that the boost operation does not have additional accompanied dilatation operation when $A_3=A_1$.

The second subfamily is denoted by $xdisim2$, in which there remain three deform parameters, $A_{1b}^1,A_{rt}^t,A_{bt}^t$, for $A_{1x}^z = A_{bt}^z = 0,A_{1x}^t = 2A_{1b}^1$, and the commutation relations become
\beq \label{eq: commrelxisim2}
\begin{array}{l}
\left[ {{t_1},{b_z}} \right] = \left( {1 + A_{1b}^1} \right){t_1}, \left[ {{t_2},{b_z}} \right] = \left( {1 + A_{1b}^1} \right){t_2},
\left[ {{t_1},{p_x}} \right] = \left( {1 + 2A_{1b}^1} \right){p_t} - {p_z}, \left[ {{t_1},{p_z}} \right] = \left( {1 + 2A_{1b}^1} \right){p_x},\\
\left[ {{t_2},{p_y}} \right] = \left( {1 + 2A_{1b}^1} \right){p_t} - {p_z}, \left[ {{t_2},{p_z}} \right] = \left( {1 + 2A_{1b}^1} \right){p_y},
\left[ {{r_z},{p_t}} \right] = A_{rt}^t{p_t}, \left[ {{r_z},{p_x}} \right] = {p_y} + A_{rt}^t{p_x},\\ \left[ {{r_z},{p_y}} \right] =  - {p_x} + A_{rt}^t{p_y}, \left[ {{r_z},{p_z}} \right] = A_{rt}^t{p_z},
\left[ {{b_z},{p_t}} \right] = {p_z} + A_{bt}^t{p_t}, \left[ {{b_z},{p_x}} \right] = \left( {A_{1b}^1 + A_{bt}^t} \right){p_x},\\
\left[ {{b_z},{p_y}} \right] = \left( {A_{1b}^1 + A_{bt}^t} \right){p_y},
\left[ {{b_z},{p_z}} \right] = \left( {1 + 2A_{1b}^1} \right){p_t} + \left( {2A_{1b}^1 + A_{bt}^t} \right){p_z}.
\end{array}
\eeq
There are many equivalent natural representations of this deformed group, one of which is as follows,
\beq \label{eq: repmatxisim2}
\begin{array}{l}
{r_z} = \left( {\begin{array}{*{20}{c}}
{{A_2}}&{}&{}&{}&{}\\
{}&{{A_2}}&{ - 1}&{}&{}\\
{}&1&{{A_2}}&{}&{}\\
{}&{}&{}&{{A_2}}&{}\\
{}&{}&{}&{}&0
\end{array}} \right),
{b_z} = \left( {\begin{array}{*{20}{c}}
{2{A_1} + {A_3}}&{}&{}&{1 + 2{A_1}}&{}\\
{}&{{A_1} + {A_3}}&{}&{}&{}\\
{}&{}&{{A_1} + {A_3}}&{}&{}\\
1&{}&{}&{{A_3}}&{}\\
{}&{}&{}&{}&0
\end{array}} \right),
{p_z} = \left( {\begin{array}{*{20}{c}}
0&{}&{}&{}&{2{A_1}}\\
{}&0&{}&{}&{}\\
{}&{}&0&{}&{}\\
{}&{}&{}&0&1\\
{}&{}&{}&{}&0
\end{array}} \right).
\end{array}
\eeq
The deform parameter in the rotation generator is supposed to be zero for the same reason that we need a resonable local rotation operation. Now  we arrive at the natural representation of the deformed single parameter group element,
\beq \label{eq: grpelexisim1}
\begin{array}{l}
{B_z}\left( \theta  \right) = {e^{\left( {{A_1} + {A_3}} \right)\theta }}\left( {\begin{array}{*{20}{c}}
{\cosh \omega  + \frac{{{A_1}}}{{1 + {A_1}}}\sinh \omega }&{}&{}&{\frac{{1 + 2{A_1}}}{{1 + {A_1}}}\sinh \omega }\\
{}&1&{}&{}\\
{}&{}&1&{}\\
{\frac{1}{{1 + {A_1}}}\sinh \omega }&{}&{}&{\cosh \omega  - \frac{{{A_1}}}{{1 + {A_1}}}\sinh \omega }
\end{array}} \right),
{P_z}\left( \lambda  \right) = \left( {\begin{array}{*{20}{c}}
{2{A_1}\lambda }\\
{}\\
{}\\
\lambda
\end{array}} \right),
\end{array}
\eeq
where $\omega  = \left( {1 + {A_1}} \right)\theta$. Note that the boost operation does not have the additional accompanied dilatation when ${A_3} =  - {A_1}$ as similar as in the previous cases.

\subsection{The deformation of $IHOM$}

The Lie algebra of semi-direct product of $HOM$ with $T(4)$ has the following commutation relations,
\[\begin{array}{l}
\left[ {{t_1},{b_z}} \right] = {t_1}, \left[ {{t_1},{p_t}} \right] = \left[ {{t_1},{p_z}} \right] = {p_x}, \left[ {{t_1},{p_x}} \right] = {p_t} - {p_z},
\left[ {{t_2},{b_z}} \right] = {t_2}, \left[ {{t_2},{p_t}} \right] = \left[ {{t_2},{p_z}} \right] = {p_y},\\
\left[ {{t_2},{p_y}} \right] = {p_t} - {p_z}, \left[ {{b_z},{p_t}} \right] = {p_z}, \left[ {{b_z},{p_z}} \right] = {p_t}
\end{array}\]
The deform group $DIHOM$ of $IHOM$ which preserves $HOM$ undeformed has four deform parameters which satisfy three second order constrain conditions,
\[A_{{\rm{1}}y}^1A_{bt}^b = A_{1y}^1A_{bt}^z = A_{2t}^b\left( {A_{bt}^t + A_{bt}^z} \right) = 0,\]
and the non-triviality condition,
\[{\left( {A_{1y}^1 + A_{2t}^b} \right)^{\rm{2}}} + {\left( {A_{bt}^t + A_{bt}^z} \right)^{\rm{2}}} \ne 0.\]
It can be classified into two families, one is denoted by $dihom1$ with $A_{1y}^1 = A_{2t}^b = 0$ and has two deform parameters $A_{bt}^t,A_{bt}^z$, the other is denoted by $dihom2$ with $A_{bt}^z =  - A_{bt}^t$ and has two deform parameters $A_{1y}^1,A_{2t}^b$.

The commutation relations for $dihom1$ is,
\beq \label{eq: commreldihom1}
\begin{array}{l}
\left[ {{b_z},{p_t}} \right] = {p_z} + A_{bt}^t{p_t} + A_{bt}^z{p_z},
\left[ {{b_z},{p_z}} \right] = {p_t} + \left( {A_{bt}^t + {\rm{2}}A_{bt}^z} \right){p_z} - A_{bt}^z{p_t},\\
\left[ {{b_z},{p_x}} \right] = \left( {A_{bt}^t + A_{bt}^z} \right){p_x}, \left[ {{b_z},{p_y}} \right] = \left( {A_{bt}^t + A_{bt}^z} \right){p_y}.
\end{array}
\eeq
Note that any value of $A_{bt}^t$ when $A_{bt}^t + A_{bt}^z$ is kept fixed gives the same deformed Lie algebra just as what happens in deformed $sim$ lie algebra. We therefore take $A_{bt}^z =0$. Note also that the commutation relation of $dihom1$ is almost the same as one of the deformed $isim$ algebra with $sim$ part invariant, the difference is that $dihom1$ has one less generators than $disim$. The deformed part of the natural representation of $dihom1$ is
\beq \label{eq: boost1dihom1}
{b_z} = \left( {\begin{array}{*{20}{c}}
{{A_1}}&{}&{}&1&{}\\
{}&{{A_1}}&{}&{}&{}\\
{}&{}&{{A_1}}&{}&{}\\
1&{}&{}&{{A_1}}&{}\\
{}&{}&{}&{}&0
\end{array}} \right),
\eeq
which is apparently the same as in $disim1$. Taking $A_{bt}^t = 0$ is another choice and the natural representation of deformation part is
\beq \label{eq: boost2dihom1}
{b_z} = \left( {\begin{array}{*{20}{c}}
0&{}&{}&{1 - {A_1}}&{}\\
{}&{{A_1}}&{}&{}&{}\\
{}&{}&{{A_1}}&{}&{}\\
{1 + {A_1}}&{}&{}&{2{A_1}}&{}\\
{}&{}&{}&{}&0
\end{array}} \right),
\eeq
which is the same as in $disim2$.

There is another deformation group $DIHOM2$ of $IHOM$, which is not isomorphic to $DIHOM1$ and its deformed part has the following commutation relations,
\beq \label{eq: commreldihom2}
\begin{array}{l}
\left[ {{t_2},{p_x}} \right] = \left( {A_{1y}^1 + A_{2t}^b} \right){t_1},
\left[ {{t_2},{p_t}} \right] = {p_y} + A_{2t}^b{b_z},\left[ {{t_2},{p_y}} \right] = {p_t} - {p_z} + \left( {2A_{1y}^1 + A_{2t}^b} \right){t_2},\left[ {{t_2},{p_z}} \right] = {p_y} + A_{2t}^b{b_z},\\
\left[ {{t_1},{p_y}} \right] = A_{1y}^1{t_1},
\left[ {{p_y},{p_t}} \right] = \left( {A_{1y}^1 + A_{2t}^b} \right){p_t} + A_{1y}^1{p_z},
\left[ {{p_y},{p_x}} \right] = \left( {A_{1y}^1 + A_{2t}^b} \right){p_x},
\left[ {{p_y},{p_z}} \right] = \left( {A_{1y}^1 + A_{2t}^b} \right){p_z} + A_{1y}^1{p_t}.
\end{array}
\eeq
The natural matrix representation therefore can be solved as,
\beq \label{eq: grpeledihom2}
\begin{array}{l}
{t_2} = \left( {\begin{array}{*{20}{c}}
{}&{}&1&{}&{}\\
{}&{}&{}&{}&{}\\
1&{}&{}&1&{}\\
{}&{}&{ - 1}&{}&{}\\
{ - \delta }&{}&{}&{ - \delta }&{}
\end{array}} \right),
{b_z} = \left( {\begin{array}{*{20}{c}}
\gamma &{}&{}&1&{}\\
{}&\gamma &{}&{}&{}\\
{}&{}&\gamma &{}&{}\\
1&{}&{}&\gamma &{}\\
{}&{}&{}&{}&\gamma
\end{array}} \right),
{p_t} = \left( {\begin{array}{*{20}{c}}
{}&{}&{}&{}&1\\
{}&{}&{}&{}&{}\\
{ - \delta }&{}&{}&{}&{}\\
{}&{}&{}&{}&{}\\
{}&{}&{}&{}&{}
\end{array}} \right),
{p_x} = \left( {\begin{array}{*{20}{c}}
{}&{}&{}&{}&{}\\
{}&{}&{}&{}&1\\
{}&\delta &{}&{}&{}\\
{}&{}&{}&{}&{}\\
{}&{}&{}&{}&{}
\end{array}} \right),\\
{p_y} = \left( {\begin{array}{*{20}{c}}
{ - \gamma {A_2}}&{}&{}&{{A_1}}&{}\\
{}&{ - \gamma {A_2}}&{}&{}&{}\\
{}&{}&{\delta  - \gamma {A_2}}&{}&1\\
{{A_1}}&{}&{}&{ - \gamma {A_2}}&{}\\
{}&{}&{}&{}&{ - \delta  - \gamma {A_2}}
\end{array}} \right),
{p_z} = \left( {\begin{array}{*{20}{c}}
{}&{}&{}&{}&{}\\
{}&{}&{}&{}&{}\\
{}&{}&{}&\delta &{}\\
{}&{}&{}&{}&1\\
{}&{}&{}&{}&{}
\end{array}} \right),
\end{array}
\eeq
where $\gamma$ is an arbitrary parameter and $\delta  = {A_1} + {A_2}$. Here we only list the matrices for deformed generators. Note that no matter what value of $\gamma$, the $(5,5)$ element of either $b_z$ or $p_y$ is nonzero. Moreover, the $5th$ row of $t_2$ is non-zero. So the matrix representation of $dihim2$ is different from ones of various deformed Lie algebra. The corresponding representation spacetime is apparently curved globally. Note also that the representation with different $\gamma$ is inequivalent in general. Take $\gamma=0$, we have
\beq \label{eq: smprepdihom2}
\begin{array}{l}
{t_2} = \left( {\begin{array}{*{20}{c}}
{}&{}&1&{}&{}\\
{}&{}&{}&{}&{}\\
1&{}&{}&1&{}\\
{}&{}&{ - 1}&{}&{}\\
{ - \delta }&{}&{}&{ - \delta }&{}
\end{array}} \right),
{p_y} = \left( {\begin{array}{*{20}{c}}
{}&{}&{}&{{A_1}}&{}\\
{}&{}&{}&{}&{}\\
{}&{}&\delta &{}&1\\
{{A_1}}&{}&{}&{}&{}\\
{}&{}&{}&{}&{ - \delta }
\end{array}} \right),
{p_t} = \left( {\begin{array}{*{20}{c}}
{}&{}&{}&{}&1\\
{}&{}&{}&{}&{}\\
{ - \delta }&{}&{}&{}&{}\\
{}&{}&{}&{}&{}\\
{}&{}&{}&{}&{}
\end{array}} \right),
{p_x} = \left( {\begin{array}{*{20}{c}}
{}&{}&{}&{}&{}\\
{}&{}&{}&{}&1\\
{}&\delta &{}&{}&{}\\
{}&{}&{}&{}&{}\\
{}&{}&{}&{}&{}
\end{array}} \right),
{p_z} = \left( {\begin{array}{*{20}{c}}
{}&{}&{}&{}&{}\\
{}&{}&{}&{}&{}\\
{}&{}&{}&\delta &{}\\
{}&{}&{}&{}&1\\
{}&{}&{}&{}&{}
\end{array}} \right).
\end{array}
\eeq
The representation of $dihom2$ is totally different from one of $dihom1$.

\subsection{The deformed group of $TE(2)$}

Just like $HOM$ group, $E(2)$ group is also the subgroup of Lorentz group with three generators. The corresponding Lie algebra is $e(2)$. The semiproduct of $E(2)$ and $T(4)$ is denoted by $TE$, and its Lie algebra is denoted by $te$ with the commutation relations,
\beq \label{eq: commrelte}
\begin{array}{l}
\left[ {{t_1},{r_z}} \right] =  - {t_2},
\left[ {{t_1},{p_t}} \right] = \left[ {{t_1},{p_z}} \right] = {p_x},
\left[ {{t_1},{p_x}} \right] = {p_t} - {p_z},
\left[ {{t_2},{r_z}} \right] = {t_1}, \\
\left[ {{t_2},{p_t}} \right] = \left[ {{t_2},{p_z}} \right] = {p_y}, \left[ {{t_2},{p_y}} \right] = {p_t} - {p_z},
\left[ {{r_z},{p_x}} \right] = {p_y}, \left[ {{r_z},{p_y}} \right] =  - {p_x}.
\end{array}
\eeq
The deformed $TE$ is $DTE$ with 5 deform parameters $A_{1t}^1,A_{rt}^t,A_{rt}^z,A_{tx}^1,A_{tx}^x$ under the second order constrain conditions,
\beq \label{eq: prmcstrte}
\left\{ \begin{array}{l}
A_{1t}^1\left( {A_{rt}^t + A_{rt}^z} \right) = 0,\\
A_{tx}^1\left( {A_{rt}^t + A_{rt}^z} \right) = 0,\\
A_{tx}^x\left( {A_{rt}^t + A_{rt}^z} \right) = 0,\\
A_{1t}^1A_{tx}^x = 0.
\end{array} \right.
\eeq
The non-triviality condition is
\beq \label{eq: ntvcdtte}
\begin{array}{c}
{\left( {A_{rt}^t + A_{rt}^z} \right)^2} + {\left( {A_{rt}^t + 2A_{rt}^z} \right)^2}
+{\left( {A_{tx}^1} \right)^2} + {\left( {A_{1t}^1 - A_{tx}^x} \right)^2} \ne 0.
\end{array}
\eeq
The $DTE$ therefore can be divided into several families similar to what happens in $DISIM$ and $DIHOM$.

\begin{enumerate}
  \item $A_{rt}^t + A_{rt}^z \ne 0,A_{1t}^1 = A_{tx}^1 = A_{tx}^x = 0$, the corresponding deformed Lie algebra is denoted by $dte1$ with the following commutation relations,
      \beq \label{eq: commreldt1}
     \begin{array}{l}
\left[ {{r_z},{p_t}} \right] = A_{rt}^t{p_t} + A_{rt}^z{p_z},
\left[ {{r_z},{p_x}} \right] = {p_y} + \left( {A_{rt}^t + A_{rt}^z} \right){p_x},\\
\left[ {{r_z},{p_y}} \right] =  - {p_x} + \left( {A_{rt}^t + A_{rt}^z} \right){p_y},
\left[ {{r_z},{p_z}} \right] = A_{rt}^z{p_t} + \left( {A_{rt}^t + 2A_{rt}^z} \right){p_z}.
\end{array}
\eeq
The non-triviality condition is now
\beq \label{eq: ntvcdtdihom2}
\begin{array}{c}
 {\left( {A_{rt}^t + A_{rt}^z} \right)^2} + {\left( {A_{rt}^t + 2A_{rt}^z} \right)^2}
 + {\left( {A_{rt}^t} \right)^2} + {\left( {A_{rt}^z} \right)^2} \ne 0.
\end{array}
\eeq

The matrix representation is
\beq \label{eq: matrepdihom2}
{r_z} = \left( {\begin{array}{*{20}{c}}
{{A_1}}&{}&{}&{ - {A_2}}&{}\\
{}&{{A_1} + {A_2}}&{ - 1}&{}&{}\\
{}&1&{{A_1} + {A_2}}&{}&{}\\
{{A_2}}&{}&{}&{{A_1} + 2{A_2}}&{}\\
{}&{}&{}&{}&0
\end{array}} \right),
\eeq
where $A_1 = A_{rt}^t$ and $A_2 = A_{rt}^z$, while the corresponding group element as
\beq \label{eq: grpeledihom2}
\begin{array}{l}
{R_z}\left( \theta  \right) = {e^{\left( {{A_1} + {A_2}} \right)\theta }}\left( {\begin{array}{*{20}{c}}
{1 - {A_2}\theta }&{}&{}&{ - {A_2}\theta }\\
{}&{\cos \theta }&{ - \sin \theta }&{}\\
{}&{\sin \theta }&{\cos \theta }&{}\\
{{A_2}\theta }&{}&{}&{1 + {A_2}\theta }
\end{array}} \right).
\end{array}
\eeq
It is apparent that the rotation operation changes a lot and may have additional accompanied dilatation as in $disim$. Moreover, the rotation itself is not only a rotation in $xy$ plane but also rotation in the rotated $tz$ plane.

  \item $A_{rt}^t + A_{rt}^z = 0,A_{1t}^1 = 0$, the corresponding Lie algebra is denoted by $dte2$. There are three deform parameters $A_{rt}^t,A_{tx}^1,A_{tx}^x$ and the commutation relations are
      \beq \label{eq:cmmtreldte2}
      \begin{array}{l}
\left[ {{r_z},{p_t}} \right] = A_{rt}^t\left( {{p_t} - {p_z}} \right),
\left[ {{r_z},{p_z}} \right] = A_{rt}^t\left( {{p_t} - {p_z}} \right),
\left[ {{p_t},{p_x}} \right] = A_{tx}^1{t_1} + A_{tx}^x{p_x},
\left[ {{p_t},{p_y}} \right] = A_{tx}^1{t_2} + A_{tx}^x{p_y},\\
 \left[ {{p_t},{p_z}} \right] = A_{tx}^x\left( {{p_z} - {p_t}} \right),
\left[ {{p_z},{p_x}} \right] = A_{tx}^1{t_1} + A_{tx}^x{p_x},
\left[ {{p_z},{p_y}} \right] = A_{tx}^1{t_2} + A_{tx}^x{p_y},
\end{array}
\eeq
which can be simplified further by a linear transformation in Lie algebra,
\beq \label{eq:trsfmatdte2}
T = \left( {\begin{array}{*{20}{c}}
1&{}&{}&{}&{}&{}&{}\\
{}&1&{}&{}&{}&{}&{}\\
{}&{}&1&\lambda &{}&{}&{ - \lambda }\\
{}&{}&{}&1&{}&{}&{}\\
{}&{}&{}&{}&1&{}&{}\\
{}&{}&{}&{}&{}&1&{}\\
{}&{}&{}&{}&{}&{}&1
\end{array}} \right).
\eeq
The new set of generators after $T$ transformation differs from the old only with $r_z$ replaced by $r{'_z} = {r_z} + \lambda \left( {{p_t} - {p_z}} \right)$. Then the new commutation relations are
\beq \label{eq: newcommreldte2}
\begin{array}{l}
\left[ {{t_1},r{'_z}} \right] = \left[ {{t_1},{r_z}} \right] + \lambda \left( {\left[ {{t_1},{p_t}} \right] - \left[ {{t_1},{p_z}} \right]} \right)
 = \left[ {{t_1},{r_z}} \right] - \lambda \left( {{p_x} - {p_x}} \right) = \left[ {{t_1},{r_z}} \right],\\
\left[ {{t_2},r{'_z}} \right] = \left[ {{t_2},{r_z}} \right] + \lambda \left( {\left[ {{t_2},{p_t}} \right] - \left[ {{t_2},{p_z}} \right]} \right)
 = \left[ {{t_2},{r_z}} \right] - \lambda \left( {{p_y} - {p_y}} \right) = \left[ {{t_2},{r_z}} \right],\\
\left[ {r{'_z},{p_t}} \right] = \left[ {{r_z},{p_t}} \right] + \lambda \left[ {{p_t},{p_z}} \right]
= A_{rt}^t\left( {{p_t} - {p_z}} \right) + \lambda A_{tx}^x\left( {{p_z} - {p_t}} \right)
= \left( {A_{rt}^t - \lambda A_{tx}^x} \right)\left( {{p_t} - {p_z}} \right),\\
\left[ {r{'_z},{p_z}} \right] = \left[ {{r_z},{p_z}} \right] + \lambda \left[ {{p_t},{p_z}} \right]
= A_{rt}^t\left( {{p_t} - {p_z}} \right) + \lambda A_{tx}^x\left( {{p_z} - {p_t}} \right)
= \left( {A_{rt}^t - \lambda A_{tx}^x} \right)\left( {{p_t} - {p_z}} \right),\\
\left[ {r{'_z},{p_x}} \right] = \left[ {{r_z},{p_x}} \right] + \lambda \left( {\left[ {{p_t},{p_x}} \right] - \left[ {{p_z},{p_x}} \right]} \right) = 0,
\left[ {r{'_z},{p_y}} \right] = \left[ {{r_z},{p_y}} \right] + \lambda \left( {\left[ {{p_t},{p_y}} \right] - \left[ {{p_z},{p_y}} \right]} \right) = 0,
\end{array}
\eeq
i.e. the commutation relations are almost kept unchanged except $\left[ {r{'_z},{p_t}} \right]$ and $\left[ {r{'_z},{p_t}} \right]$. Define $A^{\prime t}_{rt} = A_{rt}^t - \lambda A_{tx}^x$, the new commutation relations are
\beq \label{eq: newcommrel2dte2}
\left\{ \begin{array}{l}
\left[ {r{'_z},{p_t}} \right] = A^{\prime t}_{rt}\left( {{p_t} - {p_z}} \right)
= \left( {A_{rt}^t - \lambda A_{tx}^x} \right)\left( {{p_t} - {p_z}} \right),\\
\left[ {r{'_z},{p_z}} \right] = A^{\prime t}_{rt}\left( {{p_t} - {p_z}} \right)
= \left( {A_{rt}^t - \lambda A_{tx}^x} \right)\left( {{p_t} - {p_z}} \right).
\end{array} \right.
\eeq
Hence $A_{tx}^t$ and $A_{tx}^x$ are not independent parameters. We can specify two subfamily of deformation group $DTE2$ further.
One subfamily is denoted by $dte2a$ in which $A_{tx}^1,A_{tx}^x$ is taken as the independent parameters. The deformed commutation relations are
\beq \label{eq:dfmcommreldte2}
\begin{array}{l}
\left[ {{p_t},{p_x}} \right] = A_{tx}^1{t_1} + A_{tx}^x{p_x},
\left[ {{p_t},{p_y}} \right] = A_{tx}^1{t_2} + A_{tx}^x{p_y},
\left[ {{p_t},{p_z}} \right] = A_{tx}^x\left( {{p_z} - {p_t}} \right),\\
\left[ {{p_z},{p_x}} \right] = A_{tx}^1{t_1} + A_{tx}^x{p_x},
\left[ {{p_z},{p_y}} \right] = A_{tx}^1{t_2} + A_{tx}^x{p_y}.
\end{array}
\eeq
In the perturbation expansion of its matrix representation, the first order of some $\bar A_{ij}^k$ in  \eqref{eq:pertexpmatrep} do not contribute but their second order do, i.e. $A_{tx}^1 = \alpha {\tau ^2},A_{tx}^x = \beta \tau$. There are two inequivalent representations for this subfamily. One is
\beq \label{eq: matrep1dte2}
\begin{array}{l}
{p_t} = \left( {\begin{array}{*{20}{c}}
{\alpha  + 2\beta }&{}&{}&\alpha &1\\
{}&\beta &{}&{}&{}\\
{}&{}&\beta &{}&{}\\
{ - \alpha  - {A_2}}&{}&{}&{2\beta  - \alpha  - {A_2}}&{}\\
{}&{}&{}&{}&0
\end{array}} \right),
{p_x} = \left( {\begin{array}{*{20}{c}}
{}&{ - \beta }&{}&{}&{}\\
{\beta  - {A_2}}&{}&{}&{\beta  - {A_2}}&1\\
{}&{}&{}&{}&{}\\
{}&\beta &{}&{}&{}\\
{}&{}&{}&{}&0
\end{array}} \right),\\
{p_z} = \left( {\begin{array}{*{20}{c}}
{\alpha  + {A_2}}&{}&{}&{\alpha  - 2\beta  + {A_2}}&{}\\
{}&\beta &{}&{}&{}\\
{}&{}&\beta &{}&{}\\
{2\beta  - \alpha  - 2{A_2}}&{}&{}&{4\beta  - \alpha  - 2{A_2}}&1\\
{}&{}&{}&{}&0
\end{array}} \right),
{p_y} = \left( {\begin{array}{*{20}{c}}
{}&{}&{ - \beta }&{}&{}\\
{}&{}&{}&{}&{}\\
{\beta  - {A_2}}&{}&{}&{\beta  - {A_2}}&1\\
{}&{}&\beta &{}&{}\\
{}&{}&{}&{}&0
\end{array}} \right),
\end{array}
\eeq
where ${A_1} = A_{tx}^1$, ${A_2} = A_{tx}^x$, $\alpha$ is a free parameter and $\beta$ satisfies ${\beta ^2} - {A_2}\beta  + {A_1} = 0$. It can simplified by setting $\alpha  =  - \frac{{{A_2}}}{2}$
\beq \label{eq: smpfmatrep1dte2}
\begin{array}{l}
{p_t} = \left( {\begin{array}{*{20}{c}}
{2\beta  - \frac{{{A_2}}}{2}}&{}&{}&{ - \frac{{{A_2}}}{2}}&1\\
{}&\beta &{}&{}&{}\\
{}&{}&\beta &{}&{}\\
{ - \frac{{{A_2}}}{2}}&{}&{}&{2\beta  - \frac{{{A_2}}}{2}}&{}\\
{}&{}&{}&{}&0
\end{array}} \right),
{p_x} = \left( {\begin{array}{*{20}{c}}
{}&{ - \beta }&{}&{}&{}\\
{\beta  - {A_2}}&{}&{}&{\beta  - {A_2}}&1\\
{}&{}&{}&{}&{}\\
{}&\beta &{}&{}&{}\\
{}&{}&{}&{}&0
\end{array}} \right),\\
{p_z} = \left( {\begin{array}{*{20}{c}}
{\frac{{{A_2}}}{2}}&{}&{}&{\frac{{{A_2}}}{2} - 2\beta }&{}\\
{}&\beta &{}&{}&{}\\
{}&{}&\beta &{}&{}\\
{2\beta  - \frac{{3{A_2}}}{2}}&{}&{}&{4\beta  - \frac{{3{A_2}}}{2}}&1\\
{}&{}&{}&{}&0
\end{array}} \right),
{p_y} = \left( {\begin{array}{*{20}{c}}
{}&{}&{ - \beta }&{}&{}\\
{}&{}&{}&{}&{}\\
{\beta  - {A_2}}&{}&{}&{\beta  - {A_2}}&1\\
{}&{}&\beta &{}&{}\\
{}&{}&{}&{}&0
\end{array}} \right).
\end{array}
\eeq
The other kind of representation is
\beq \label{eq: matrep2dte2}
\begin{array}{l}
{p_t} = \left( {\begin{array}{*{20}{c}}
\gamma &{}&{}&{\gamma  - {A_2}}&1\\
{}&\lambda &{}&{}&{}\\
{}&{}&\lambda &{}&{}\\
{2\lambda  - \gamma  - {A_2}}&{}&{}&{2\lambda  - \gamma }&{}\\
{}&{}&{}&{}&0
\end{array}} \right),
{p_x} = \left( {\begin{array}{*{20}{c}}
{}&{\lambda  - {A_2}}&{}&{}&{}\\
{\lambda  - {A_2}}&{}&{}&{\lambda  - {A_2}}&1\\
{}&{}&{}&{}&{}\\
{}&{{A_2} - \lambda }&{}&{}&{}\\
{}&{}&{}&{}&0
\end{array}} \right),\\
{p_y} = \left( {\begin{array}{*{20}{c}}
{}&{}&{\lambda  - {A_2}}&{}&{}\\
{}&{}&{}&{}&{}\\
{\lambda  - {A_2}}&{}&{}&{\lambda  - {A_2}}&1\\
{}&{}&{{A_2} - \lambda }&{}&{}\\
{}&{}&{}&{}&0
\end{array}} \right),
{p_z} = \left( {\begin{array}{*{20}{c}}
\gamma &{}&{}&{\gamma  - {A_2}}&{}\\
{}&\lambda &{}&{}&{}\\
{}&{}&\lambda &{}&{}\\
{2\lambda  - \gamma  - {A_2}}&{}&{}&{2\lambda  - \gamma }&{}\\
{}&{}&{}&{}&0
\end{array}} \right),

\end{array}
\eeq
where $\gamma$ is a free parameter and $\lambda$ satisfies ${\lambda ^2} - {A_2}\lambda  + {A_1} = 0$. Setting $\gamma =\lambda$, it is simplified as
\beq \label{eq: smpfmatrep2dte2}
\begin{array}{l}
{p_t} = \left( {\begin{array}{*{20}{c}}
\lambda &{}&{}&{\lambda  - {A_2}}&1\\
{}&\lambda &{}&{}&{}\\
{}&{}&\lambda &{}&{}\\
{\lambda  - {A_2}}&{}&{}&\lambda &{}\\
{}&{}&{}&{}&0
\end{array}} \right),
{p_x} = \left( {\begin{array}{*{20}{c}}
{}&{\lambda  - {A_2}}&{}&{}&{}\\
{\lambda  - {A_2}}&{}&{}&{\lambda  - {A_2}}&1\\
{}&{}&{}&{}&{}\\
{}&{{A_2} - \lambda }&{}&{}&{}\\
{}&{}&{}&{}&0
\end{array}} \right),\\
{p_y} = \left( {\begin{array}{*{20}{c}}
{}&{}&{\lambda  - {A_2}}&{}&{}\\
{}&{}&{}&{}&{}\\
{\lambda  - {A_2}}&{}&{}&{\lambda  - {A_2}}&1\\
{}&{}&{{A_2} - \lambda }&{}&{}\\
{}&{}&{}&{}&0
\end{array}} \right).
{p_z} = \left( {\begin{array}{*{20}{c}}
\lambda &{}&{}&{\lambda  - {A_2}}&{}\\
{}&\lambda &{}&{}&{}\\
{}&{}&\lambda &{}&{}\\
{\lambda  - {A_2}}&{}&{}&\lambda &{}\\
{}&{}&{}&{}&0
\end{array}} \right),
\end{array}
\eeq
It is apparent that the translation operations are entangled with $t_1$ and $t_2$ operations together in both representations.

The other subfamily is denoted by $dte2b$ in which $A^1_{tx}$ and $A^t_{rt}$ is taken as independent deform parameters. Its commutation relation is
\beq \label{eq:commreldte2b}
\begin{array}{l}
\left[ {{r_z},{p_t}} \right] = A_{rt}^t\left( {{p_t} - {p_z}} \right),
\left[ {{r_z},{p_z}} \right] = A_{rt}^t\left( {{p_t} - {p_z}} \right),\\
\left[ {{p_t},{p_x}} \right] = A_{tx}^1{t_1}, \left[ {{p_t},{p_y}} \right] = A_{tx}^1{t_2},
\left[ {{p_z},{p_x}} \right] = A_{tx}^1{t_1}, \left[ {{p_z},{p_y}} \right] = A_{tx}^1{t_2}.
\end{array}
\eeq
$dte2b$ does not have a natural representation which is a continuous deformation from the representation of Poincar\'e group. Moreover we can observe that the deformation group is more likely an isometry group of curved spacetime and the rotation operation does not seem compact anymore. We can ignore this kind of deformation group of $E(2)$.
  \item $A_{rt}^t + A_{rt}^z = 0,A_{tx}^x = 0$ and the corresponding Lie algebra is denoted by $dte3$ with three deform parameters, $A_{1t}^1,A_{rt}^t$ and $A_{tx}^1$. The commutation relations is
      \beq \label{eq: commreldte3}
\begin{array}{l}
\left[ {{t_1},{p_t}} \right] = {p_x} + A_{1t}^1{t_1},
\left[ {{t_1},{p_z}} \right] = {p_x} + A_{1t}^1{t_1},
\left[ {{r_z},{p_t}} \right] = \left[ {{r_z},{p_z}} \right] = A_{rt}^t\left( {{p_t} - {p_z}} \right),
\left[ {{r_z},{p_x}} \right] = {p_y} - A_{1t}^1{t_2}, \\
\left[ {{r_z},{p_y}} \right] =  - {p_x} - A_{1t}^1{t_1},
\left[ {{p_t},{p_z}} \right] = A_{1t}^1\left( {{p_t} - {p_z}} \right),
\left[ {{p_t},{p_x}} \right] = \left[ {{p_z},{p_x}} \right] = A_{tx}^1{t_1},
\left[ {{p_t},{p_y}} \right] = \left[ {{p_z},{p_y}} \right] = A_{tx}^1{t_2} - A_{1t}^1{p_y}.
\end{array}
\eeq
As in the case of $dte2$, $dte3$ can be specified into two subfamilies for there is only one independent parameter from $A_{rt}^t$ and $A_{1t}^1$ via the linear combination between generators when $A_{1t}^1 \ne 0$.

The first subfamily is denoted by $det3a$, in which we take $A_{1t}^1,A_{tx}^1$ as deform parameters and the deformed commutation relations are
\beq \label{eq: commreldte3a}
\begin{array}{l}
\left[ {{t_1},{p_t}} \right] = {p_x} + A_{1t}^1{t_1},
\left[ {{t_1},{p_z}} \right] = {p_x} + A_{1t}^1{t_1},
\left[ {{r_z},{p_x}} \right] = {p_y} - A_{1t}^1{t_2},
\left[ {{r_z},{p_y}} \right] =  - {p_x} - A_{1t}^1{t_1},\\
\left[ {{p_t},{p_x}} \right] = \left[ {{p_z},{p_x}} \right] = A_{tx}^1{t_1},
\left[ {{p_t},{p_z}} \right] = A_{1t}^1\left( {{p_t} - {p_z}} \right),
\left[ {{p_t},{p_y}} \right] = \left[ {{p_z},{p_y}} \right] = A_{tx}^1{t_2} - A_{1t}^1{p_y}.
\end{array}
\eeq
Like what encounters in $dte2$, the first order of some $\bar A_{ij}^k$ in  \eqref{eq:pertexpmatrep} do not contribute but their second order do in the perturbation expansion of its matrix representation. There are two inequivalent representations for this kind. The first one is
\beq \label{eq:matrep1dte3a}
\begin{array}{l}
{p_t} = \left( {\begin{array}{*{20}{c}}
{2\alpha  + \beta }&{}&{}&\beta &1\\
{}&\alpha &{}&{}&{}\\
{}&{}&\alpha &{}&{}\\
{{A_1} - \beta }&{}&{}&{{A_1} + 2\alpha  - \beta }&{}\\
{}&{}&{}&{}&0
\end{array}} \right),
{p_x} = \left( {\begin{array}{*{20}{c}}
{}&{ - {A_1} - \alpha }&{}&{}&{}\\
\alpha &{}&{}&\alpha &1\\
{}&{}&{}&{}&{}\\
{}&{{A_1} + \alpha }&{}&{}&{}\\
{}&{}&{}&{}&0
\end{array}} \right),\\
{p_y} = \left( {\begin{array}{*{20}{c}}
{}&{}&{ - \alpha }&{}&{}\\
{}&{}&{}&{}&{}\\
{{A_1} + \alpha }&{}&{}&{{A_1} + \alpha }&1\\
{}&{}&\alpha &{}&{}\\
{}&{}&{}&{}&0
\end{array}} \right),
{p_z} = \left( {\begin{array}{*{20}{c}}
{\beta  - {A_1}}&{}&{}&{\beta  - {A_1} - 2\alpha }&{}\\
{}&\alpha &{}&{}&{}\\
{}&{}&\alpha &{}&{}\\
{2{A_1} - \beta  + 2\alpha }&{}&{}&{2{A_1} - \beta  + 4\alpha }&1\\
{}&{}&{}&{}&0
\end{array}} \right),

\end{array}
\eeq
where $\beta$ is a free parameter and $\alpha$ satisfies ${A_2} + \alpha \left( {{A_1} + \alpha } \right) = 0$. By taking $\beta  = \frac{{{A_1}}}{2}$, it is simplified to
\beq \label{eq:smpfmatrep1dte3a}
\begin{array}{l}
{p_t} = \left( {\begin{array}{*{20}{c}}
{2\alpha  + \frac{{{A_1}}}{2}}&{}&{}&{\frac{{{A_1}}}{2}}&1\\
{}&\alpha &{}&{}&{}\\
{}&{}&\alpha &{}&{}\\
{\frac{{{A_1}}}{2}}&{}&{}&{2\alpha  + \frac{{{A_1}}}{2}}&{}\\
{}&{}&{}&{}&0
\end{array}} \right),
{p_x} = \left( {\begin{array}{*{20}{c}}
{}&{ - {A_1} - \alpha }&{}&{}&{}\\
\alpha &{}&{}&\alpha &1\\
{}&{}&{}&{}&{}\\
{}&{{A_1} + \alpha }&{}&{}&{}\\
{}&{}&{}&{}&0
\end{array}} \right),\\
{p_y} = \left( {\begin{array}{*{20}{c}}
{}&{}&{ - \alpha }&{}&{}\\
{}&{}&{}&{}&{}\\
{{A_1} + \alpha }&{}&{}&{{A_1} + \alpha }&1\\
{}&{}&\alpha &{}&{}\\
{}&{}&{}&{}&0
\end{array}} \right).
{p_z} = \left( {\begin{array}{*{20}{c}}
{ - \frac{{{A_1}}}{2}}&{}&{}&{ - 2\alpha  - \frac{{{A_1}}}{2}}&{}\\
{}&\alpha &{}&{}&{}\\
{}&{}&\alpha &{}&{}\\
{2\alpha  + \frac{{3{A_1}}}{2}}&{}&{}&{4\alpha  + \frac{{3{A_1}}}{2}}&1\\
{}&{}&{}&{}&0
\end{array}} \right),

\end{array}
\eeq
The second representation is
\beq \label{eq:matrep2dte3a}
\begin{array}{l}
{p_t} = \left( {\begin{array}{*{20}{c}}
\lambda &{}&{}&{{A_1} + \lambda }&1\\
{}&\gamma &{}&{}&{}\\
{}&{}&\gamma &{}&{}\\
{{A_1} + 2\gamma  - \lambda }&{}&{}&{2\gamma  - \lambda }&{}\\
{}&{}&{}&{}&0
\end{array}} \right),
{p_x} = \left( {\begin{array}{*{20}{c}}
{}&\gamma &{}&{}&{}\\
\gamma &{}&{}&\gamma &1\\
{}&{}&{}&{}&{}\\
{}&{ - \gamma }&{}&{}&{}\\
{}&{}&{}&{}&0
\end{array}} \right),\\
{p_y} = \left( {\begin{array}{*{20}{c}}
{}&{}&{{A_1} + \gamma }&{}&{}\\
{}&{}&{}&{}&{}\\
{{A_1} + \gamma }&{}&{}&{{A_1} + \gamma }&1\\
{}&{}&{ - {A_1} - \gamma }&{}&{}\\
{}&{}&{}&{}&0
\end{array}} \right),
{p_z} = \left( {\begin{array}{*{20}{c}}
\lambda &{}&{}&{{A_1} + \lambda }&{}\\
{}&\gamma &{}&{}&{}\\
{}&{}&\gamma &{}&{}\\
{{A_1} + 2\gamma  - \lambda }&{}&{}&{2\gamma  - \lambda }&1\\
{}&{}&{}&{}&0
\end{array}} \right),

\end{array}
\eeq
where $\lambda$ is a free parameter and $\gamma$ satisfies ${A_2} + \gamma \left( {{A_1} + \gamma } \right) = 0$.
By taking $\lambda =\gamma$, it is simplified to
\beq \label{eq:smpfmatrep2dte3a}
\begin{array}{l}
{p_t} = \left( {\begin{array}{*{20}{c}}
\gamma &{}&{}&{{A_1} + \gamma }&1\\
{}&\gamma &{}&{}&{}\\
{}&{}&\gamma &{}&{}\\
{{A_1} + \gamma }&{}&{}&\gamma &{}\\
{}&{}&{}&{}&0
\end{array}} \right),
{p_x} = \left( {\begin{array}{*{20}{c}}
{}&\gamma &{}&{}&{}\\
\gamma &{}&{}&\gamma &1\\
{}&{}&{}&{}&{}\\
{}&{ - \gamma }&{}&{}&{}\\
{}&{}&{}&{}&0
\end{array}} \right),\\
{p_y} = \left( {\begin{array}{*{20}{c}}
{}&{}&{{A_1} + \gamma }&{}&{}\\
{}&{}&{}&{}&{}\\
{{A_1} + \gamma }&{}&{}&{{A_1} + \gamma }&1\\
{}&{}&{ - {A_1} - \gamma }&{}&{}\\
{}&{}&{}&{}&0
\end{array}} \right),
{p_z} = \left( {\begin{array}{*{20}{c}}
\gamma &{}&{}&{{A_1} + \gamma }&{}\\
{}&\gamma &{}&{}&{}\\
{}&{}&\gamma &{}&{}\\
{{A_1} + \gamma }&{}&{}&\gamma &1\\
{}&{}&{}&{}&0
\end{array}} \right).

\end{array}
\eeq
It is apparent again as in the $DTE2a$ that the translation operations are entangled with $t_1$ and $t_2$ operations together in both representations.

The second subfamily is denoted by $dte3b$, in which we take $A_{rt}^t,A_{tx}^1$ as deform parameters and the deformed commutation relations are
\beq \label{eq:commreldte3b}
\begin{array}{l}
\left[ {{r_z},{p_t}} \right] = \left[ {{r_z},{p_z}} \right] = A_{rt}^t\left( {{p_t} - {p_z}} \right),
\left[ {{p_t},{p_x}} \right] = \left[ {{p_z},{p_x}} \right] = A_{tx}^1{t_1},
\left[ {{p_t},{p_y}} \right] = \left[ {{p_z},{p_y}} \right] = A_{tx}^1{t_2}.
\end{array}
\eeq
the corresponding deformed matrix representation is
\beq \label{eq:matrepdte3b}
\begin{array}{l}
{p_t} = \left( {\begin{array}{*{20}{c}}
{}&{}&{}&{{A_1}}&1\\
{}&{ - {A_1}}&{}&{}&{}\\
{}&{}&{ - {A_1}}&{}&{}\\
{ - {A_1}}&{}&{}&{ - 2{A_1}}&{}\\
{}&{}&{}&{}&0
\end{array}} \right),
{r_z} = \left( {\begin{array}{*{20}{c}}
{{A_2}}&{}&{}&{{A_2}}&{}\\
{}&{}&{ - 1}&{}&{}\\
{}&1&{}&{}&{}\\
{ - {A_2}}&{}&{}&{ - {A_2}}&{}\\
{}&{}&{}&{}&0
\end{array}} \right),\\
{p_x} = \left( {\begin{array}{*{20}{c}}
{}&{ - {A_1}}&{}&{}&{}\\
{ - {A_1}}&{}&{}&{ - {A_1}}&1\\
{}&{}&{}&{}&{}\\
{}&{{A_1}}&{}&{}&{}\\
{}&{}&{}&{}&0
\end{array}} \right),
{p_z} = \left( {\begin{array}{*{20}{c}}
{}&{}&{}&{{A_1}}&{}\\
{}&{ - {A_1}}&{}&{}&{}\\
{}&{}&{ - {A_1}}&{}&{}\\
{ - {A_1}}&{}&{}&{ - 2{A_1}}&1\\
{}&{}&{}&{}&0
\end{array}} \right),

\end{array}
\eeq
where the sinle parameter group element representation corresponding to $r_z$ is

\beq \label{eq:grpeledte3b}
{R_z}\left( \theta  \right) = \left( {\begin{array}{*{20}{c}}
{1 + {A_2}\theta }&{}&{}&{{A_2}\theta }\\
{}&{\cos \theta }&{ - \sin \theta }&{}\\
{}&{\sin \theta }&{\cos \theta }&{}\\
{ - {A_2}\theta }&{}&{}&{1 - {A_2}\theta }
\end{array}} \right),
\eeq
a reasonable rotation operation not only in $xy$ plane but also in rotated $tz$ plane as in $DTE1$. The translation operations are entangled with $t_1$ and $t_2$ operations together again.
\end{enumerate}

The common feathers of $DTE$ are that the rotation operation is not only in $xy$ plane but also in rotated $tz$ plane and the translation operations are entangled with $t_1$ and $t_2$ operations together.

\subsection{The deformation group of $ISO(3)$}

$SO(3)$ group has three generators ${r_x},{r_y},{r_z}$. The deformation of its semi-direct product with $T(4)$ has two deform parameters $A_{tx}^x,A_{xy}^3$, where 3 represents $r_z$. The second order constrain condition is
\beq \label{eq:so3secord}
A_{tx}^xA_{xy}^3 = 0
\eeq
The deformation group $DISO(3)$ therefore canbe specified into two classes.
The first class is denoted by $diso(3)1$, in which the deform parameter is taken as $A_1 =A_{xy}^3$ and the commutation relations are
\beq \label{eq:diso31commrel}
\left[ {{p_x},{p_y}} \right] = {A_1}{r_z}, \left[ {{p_z},{p_x}} \right] = {A_1}{r_y}, \left[ {{p_y},{p_z}} \right] = {A_1}{r_x}.
\eeq
The natural matrix representation is
\beq \label{eq:diso31rep}
\begin{array}{l}
{p_t} = \left( {\begin{array}{*{20}{c}}
\alpha &{}&{}&{}&1\\
{}&\alpha &{}&{}&{}\\
{}&{}&\alpha &{}&{}\\
{}&{}&{}&\alpha &{}\\
{}&{}&{}&{}&0
\end{array}} \right),
{p_x} = \left( {\begin{array}{*{20}{c}}
{}&\beta &{}&{}&{}\\
\alpha &{}&{}&{}&1\\
{}&{}&0&{}&{}\\
{}&{}&{}&0&{}\\
{}&{}&{}&{}&0
\end{array}} \right),
{p_y} = \left( {\begin{array}{*{20}{c}}
{}&{}&\beta &{}&{}\\
{}&0&{}&{}&{}\\
\alpha &{}&{}&{}&1\\
{}&{}&{}&0&{}\\
{}&{}&{}&{}&0
\end{array}} \right),{p_z} = \left( {\begin{array}{*{20}{c}}
{}&{}&{}&\beta &{}\\
{}&0&{}&{}&{}\\
{}&{}&0&{}&{}\\
\alpha&{}&{}&{}&1\\
{}&{}&{}&{}&0
\end{array}} \right),
\end{array}
\eeq
where $\alpha$ and $\beta$ satisfy $\alpha \beta  + {A_1} = 0$. Hence there are two ways to get simplification.

In the first way, by taking $\beta=\alpha$ if ${A_1} < 0$, we have
\beq \label{eq:diso31rep1}
\begin{array}{l}
{p_t} = \left( {\begin{array}{*{20}{c}}
\alpha &{}&{}&{}&1\\
{}&\alpha &{}&{}&{}\\
{}&{}&\alpha &{}&{}\\
{}&{}&{}&\alpha &{}\\
{}&{}&{}&{}&0
\end{array}} \right),
{p_x} = \left( {\begin{array}{*{20}{c}}
{}&\alpha &{}&{}&{}\\
\alpha &{}&{}&{}&1\\
{}&{}&0&{}&{}\\
{}&{}&{}&0&{}\\
{}&{}&{}&{}&0
\end{array}} \right),
{p_y} = \left( {\begin{array}{*{20}{c}}
{}&{}&\alpha &{}&{}\\
{}&0&{}&{}&{}\\
\alpha &{}&{}&{}&1\\
{}&{}&{}&0&{}\\
{}&{}&{}&{}&0
\end{array}} \right), {p_z} = \left( {\begin{array}{*{20}{c}}
{}&{}&{}&\alpha &{}\\
{}&0&{}&{}&{}\\
{}&{}&0&{}&{}\\
\alpha&{}&{}&{}&1\\
{}&{}&{}&{}&0
\end{array}} \right),
\end{array}
\eeq
where ${\alpha ^2} =  - {A_1}$.

In the second way, by taking $\beta=-\alpha$ if ${A_1} > 0$, we have
\beq \label{eq:diso31rep2}
\begin{array}{l}
{p_t} = \left( {\begin{array}{*{20}{c}}
\alpha &{}&{}&{}&1\\
{}&\alpha &{}&{}&{}\\
{}&{}&\alpha &{}&{}\\
{}&{}&{}&\alpha &{}\\
{}&{}&{}&{}&0
\end{array}} \right),
{p_x} = \left( {\begin{array}{*{20}{c}}
{}&-\alpha &{}&{}&{}\\
\alpha &{}&{}&{}&1\\
{}&{}&0&{}&{}\\
{}&{}&{}&0&{}\\
{}&{}&{}&{}&0
\end{array}} \right),
{p_y} = \left( {\begin{array}{*{20}{c}}
{}&{}&-\alpha &{}&{}\\
{}&0&{}&{}&{}\\
\alpha &{}&{}&{}&1\\
{}&{}&{}&0&{}\\
{}&{}&{}&{}&0
\end{array}} \right),
{p_z} = \left( {\begin{array}{*{20}{c}}
{}&{}&{}&-\alpha &{}\\
{}&0&{}&{}&{}\\
{}&{}&0&{}&{}\\
\alpha&{}&{}&{}&1\\
{}&{}&{}&{}&0
\end{array}} \right),
\end{array}
\eeq
where ${\alpha ^2} ={A_1}$.

The second family of deformation, denoted by $diso(3)2$, consists of deformation with deform parameters $A_1= A_{tx}^x$. The deformed commutation relations are
\beq \label{eq:diso32commrel}
\left[ {{p_t},{p_i}} \right] = A_{tx}^x{p_i}, i = x,y,z.
\eeq

There are three kind of representation therefore. The deformed representation matrices are
\begin{enumerate}
  \item \beq \label{eq:diso32rep1}
  {p_t} = \left( {\begin{array}{*{20}{c}}
\alpha &{}&{}&{}&1\\
{}&{{A_1}}&{}&{}&{}\\
{}&{}&{{A_1}}&{}&{}\\
{}&{}&{}&{{A_1}}&{}\\
{}&{}&{}&{}&0
\end{array}} \right),
\eeq
  \item \beq \label{eq:diso32rep2}
  \begin{array}{l}
{p_t} = \left( {\begin{array}{*{20}{c}}
{ - {A_1}}&{}&{}&{}&1\\
{}&0&{}&{}&{}\\
{}&{}&0&{}&{}\\
{}&{}&{}&0&{}\\
{}&{}&{}&{}&0
\end{array}} \right),
{p_x} = \left( {\begin{array}{*{20}{c}}
0&{}&{}&{}&{}\\
{ - {A_1}}&{}&{}&{}&1\\
{}&{}&0&{}&{}\\
{}&{}&{}&0&{}\\
{}&{}&{}&{}&0
\end{array}} \right),
{p_y} = \left( {\begin{array}{*{20}{c}}
0&{}&{}&{}&{}\\
{}&0&{}&{}&{}\\
{ - {A_1}}&{}&{}&{}&1\\
{}&{}&{}&0&{}\\
{}&{}&{}&{}&0
\end{array}} \right),
{p_z} = \left( {\begin{array}{*{20}{c}}
0&{}&{}&{}&{}\\
{}&0&{}&{}&{}\\
{}&{}&0&{}&{}\\
{ - {A_1}}&{}&{}&{}&1\\
{}&{}&{}&{}&0
\end{array}} \right),
\end{array}
\eeq
  \item \beq \label{eq:diso32rep3}
  \begin{array}{l}
{p_t} = \left( {\begin{array}{*{20}{c}}
{2{A_1}}&{}&{}&{}&1\\
{}&{{A_1}}&{}&{}&{}\\
{}&{}&{{A_1}}&{}&{}\\
{}&{}&{}&{{A_1}}&{}\\
{}&{}&{}&{}&0
\end{array}} \right),
{p_x} = \left( {\begin{array}{*{20}{c}}
0&\alpha &{}&{}&{}\\
{}&0&{}&{}&1\\
{}&{}&0&{}&{}\\
{}&{}&{}&0&{}\\
{}&{}&{}&{}&0
\end{array}} \right),
{p_y} = \left( {\begin{array}{*{20}{c}}
0&{}&\alpha &{}&{}\\
{}&0&{}&{}&{}\\
{}&{}&0&{}&1\\
{}&{}&{}&0&{}\\
{}&{}&{}&{}&0
\end{array}} \right),
{p_z} = \left( {\begin{array}{*{20}{c}}
0&{}&{}&\alpha &{}\\
{}&0&{}&{}&{}\\
{}&{}&0&{}&{}\\
{}&{}&{}&0&1\\
{}&{}&{}&{}&0
\end{array}} \right),
\end{array}
\eeq
where $\alpha$ is a free parameter and can be taken as $A_1$ in all of the three cases.
\end{enumerate}

\subsection{The deformation of $ISO(2,1)$}

Let us investigate the deformation of semi-product of three generators Lorentz subgroup $SO(2,1)$ with $T(4)$, $DISO(2,1)$, at last. The three generators of $SO(2,1)$ are ${r_x},{b_y}$ and ${b_z}$. $DISO(2,1)$ has two deform parameters $A_{tx}^t$ and $A_{ty}^2$, where 2 represents ${b_y}$, and a second order constrain condition,
\beq \label{eq:diso21cstr}
A_{tx}^tA_{ty}^2 = 0.
\eeq
Thus $DISO(2,1)1$ can be specified into two families.

The first family is denoted by $diso(2,1)$, in which the deform parameter is taken as $A_1 =A_{ty}^2$ and the deformed commutation relations are
\beq \label{eq:diso211commrel}
\left[ {{p_t},{p_y}} \right] = {A_1}{b_y}, \left[ {{p_t},{p_z}} \right] = {A_1}{b_z}, \left[ {{p_y},{p_z}} \right] =  - {A_1}{r_x},
\eeq
as well as the representation is
\beq \label{eq:diso211rep}
\begin{array}{l}
{p_x} = \left( {\begin{array}{*{20}{c}}
\alpha &{}&{}&{}&{}\\
{}&\alpha &{}&{}&1\\
{}&{}&\alpha &{}&{}\\
{}&{}&{}&\alpha &{}\\
{}&{}&{}&{}&0
\end{array}} \right),
{p_t} = \left( {\begin{array}{*{20}{c}}
{}&\alpha &{}&{}&1\\
\beta &{}&{}&{}&{}\\
{}&{}&0&{}&{}\\
{}&{}&{}&0&{}\\
{}&{}&{}&{}&0
\end{array}} \right),
{p_y} = \left( {\begin{array}{*{20}{c}}
0&{}&{}&{}&{}\\
{}&{}&{ - \beta }&{}&{}\\
{}&\alpha &{}&{}&1\\
{}&{}&{}&0&{}\\
{}&{}&{}&{}&0
\end{array}} \right),
{p_z} = \left( {\begin{array}{*{20}{c}}
0&{}&{}&{}&{}\\
{}&{}&{}&{ - \beta }&{}\\
{}&{}&0&{}&{}\\
{}&\alpha &{}&{}&1\\
{}&{}&{}&{}&0
\end{array}} \right),
\end{array}
\eeq
where  $\alpha$ and $\beta$ satisfy $\alpha \beta  + {A_1} = 0$. Thus it can be simplified according to value of $A_1$.

When $A_1>0$, we can take $\beta =-\alpha$ and get
\beq \label{eq:diso211rep1}
\begin{array}{l}
{p_x} = \left( {\begin{array}{*{20}{c}}
\alpha &{}&{}&{}&{}\\
{}&\alpha &{}&{}&1\\
{}&{}&\alpha &{}&{}\\
{}&{}&{}&\alpha &{}\\
{}&{}&{}&{}&0
\end{array}} \right),
{p_t} = \left( {\begin{array}{*{20}{c}}
{}&\alpha &{}&{}&1\\
-\alpha &{}&{}&{}&{}\\
{}&{}&0&{}&{}\\
{}&{}&{}&0&{}\\
{}&{}&{}&{}&0
\end{array}} \right),
{p_y} = \left( {\begin{array}{*{20}{c}}
0&{}&{}&{}&{}\\
{}&{}&{ \alpha }&{}&{}\\
{}&\alpha &{}&{}&1\\
{}&{}&{}&0&{}\\
{}&{}&{}&{}&0
\end{array}} \right),
{p_z} = \left( {\begin{array}{*{20}{c}}
0&{}&{}&{}&{}\\
{}&{}&{}&{ \alpha }&{}\\
{}&{}&0&{}&{}\\
{}&\alpha &{}&{}&1\\
{}&{}&{}&{}&0
\end{array}} \right),
\end{array}
\eeq
where $\alpha  =  \pm \sqrt {{A_1}}$.

When $A_1<0$, we can take $\beta =\alpha$ and get
\beq \label{eq:diso211rep2}
\begin{array}{l}
{p_x} = \left( {\begin{array}{*{20}{c}}
\alpha &{}&{}&{}&{}\\
{}&\alpha &{}&{}&1\\
{}&{}&\alpha &{}&{}\\
{}&{}&{}&\alpha &{}\\
{}&{}&{}&{}&0
\end{array}} \right),
{p_t} = \left( {\begin{array}{*{20}{c}}
{}&\alpha &{}&{}&1\\
\alpha &{}&{}&{}&{}\\
{}&{}&0&{}&{}\\
{}&{}&{}&0&{}\\
{}&{}&{}&{}&0
\end{array}} \right),
{p_y} = \left( {\begin{array}{*{20}{c}}
0&{}&{}&{}&{}\\
{}&{}&{ -\alpha }&{}&{}\\
{}&\alpha &{}&{}&1\\
{}&{}&{}&0&{}\\
{}&{}&{}&{}&0
\end{array}} \right),
{p_z} = \left( {\begin{array}{*{20}{c}}
0&{}&{}&{}&{}\\
{}&{}&{}&{ -\alpha }&{}\\
{}&{}&0&{}&{}\\
{}&\alpha &{}&{}&1\\
{}&{}&{}&{}&0
\end{array}} \right),
\end{array}
\eeq
where $\alpha  =  \pm \sqrt {{-A_1}}$.

The second family is denoted by $diso(2,1)2$, in which the deform parameter is taken as $A_1 =A_{tx}^t$ and the deformed commutation relations are
\beq \label{eq:diso21commrel}
\left[ {{p_x},{p_i}} \right] =  - {A_1}{p_i}, i = t,y,z,
\eeq
as well as the representation is
\beq \label{eq:diso212rep}
\begin{array}{l}
{p_x} = \left( {\begin{array}{*{20}{c}}
{\alpha  - {A_1}}&{}&{}&{}&{}\\
{}&\beta &{}&{}&1\\
{}&{}&{\alpha  - {A_1}}&{}&{}\\
{}&{}&{}&{\alpha  - {A_1}}&{}\\
{}&{}&{}&{}&0
\end{array}} \right),
{p_t} = \left( {\begin{array}{*{20}{c}}
0&\alpha &{}&{}&1\\
{}&0&{}&{}&{}\\
{}&{}&0&{}&{}\\
{}&{}&{}&0&{}\\
{}&{}&{}&{}&0
\end{array}} \right),
{p_y} = \left( {\begin{array}{*{20}{c}}
0&{}&{}&{}&{}\\
{}&0&{}&{}&{}\\
{}&\alpha &0&{}&1\\
{}&{}&{}&0&{}\\
{}&{}&{}&{}&0
\end{array}} \right),
{p_z} = \left( {\begin{array}{*{20}{c}}
0&{}&{}&{}&{}\\
{}&0&{}&{}&{}\\
{}&{}&0&{}&{}\\
{}&\alpha &{}&0&1\\
{}&{}&{}&{}&0
\end{array}} \right),
\end{array}
\eeq
where $\alpha$ and $\beta$ satisfy $\alpha(\alpha-\beta) = 0$. Thus it can be simplified in two ways.

In the first way, $\alpha =0$ and hence only the representation of $p_x$ is deformed,
\beq \label{eq:diso212rep1}
{p_x} = \left( {\begin{array}{*{20}{c}}
{ - {A_1}}&{}&{}&{}&{}\\
{}&{ - {A_1}}&{}&{}&1\\
{}&{}&{ - {A_1}}&{}&{}\\
{}&{}&{}&{ - {A_1}}&{}\\
{}&{}&{}&{}&0
\end{array}} \right).
\eeq
In the second way, we take $\alpha =\beta =A_1$, the representation is simplified as
\beq \label{eq:diso212rep2}
\begin{array}{l}
{p_t} = \left( {\begin{array}{*{20}{c}}
0&{{A_1}}&{}&{}&1\\
{}&0&{}&{}&{}\\
{}&{}&0&{}&{}\\
{}&{}&{}&0&{}\\
{}&{}&{}&{}&0
\end{array}} \right),
{p_x} = \left( {\begin{array}{*{20}{c}}
0&{}&{}&{}&{}\\
{}&{{A_1}}&{}&{}&1\\
{}&{}&0&{}&{}\\
{}&{}&{}&0&{}\\
{}&{}&{}&{}&0
\end{array}} \right),
{p_y} = \left( {\begin{array}{*{20}{c}}
0&{}&{}&{}&{}\\
{}&0&{}&{}&{}\\
{}&{{A_1}}&0&{}&1\\
{}&{}&{}&0&{}\\
{}&{}&{}&{}&0
\end{array}} \right),
{p_z} = \left( {\begin{array}{*{20}{c}}
0&{}&{}&{}&{}\\
{}&0&{}&{}&{}\\
{}&{}&0&{}&{}\\
{}&{{A_1}}&{}&0&1\\
{}&{}&{}&{}&0
\end{array}} \right).
\end{array}
\eeq

\subsection{Summary, Conclusion and Outlook}

Now we investigate the deformation of semi-product of all of three and four generators Lorentz subgroups with $T(4)$ and obtain their natural representations. We list the deformation classification and the brief remark on their characters and their natural representations in Table I.

\begin{table}
\begin{center}
\caption{The Deformation of Semi-product Poincar\'e Subgroups.}
\begin{tabular}{c|c|c|c|c}
\hline
\hline
\multirow{2}{*}{subgroup} & deformation & deformation & natural & \multirow{2}{*}{remark}\\
 & family & subfamily& rep. & \\
  \hline
  \hline
  Poincar\'e  & de Sitter & de Sitter & 1 & the isometry group of maximal symmetric space of 4-spacetime\\
  \hline

\multirow{6}{*}{$ISIM$} &
{$DISIM$}&
\multirow{2}{*}{$DISIM$}&
\multirow{2}{*}{1}&
{lots of equivalent deformation corresponding to generators redefinition}\\
&($SIM$ undeformed)& & & additional accompanied dilatation for rotation and boost operation\\
\cline{2-5}
&
$XDISIM1$&
\multirow{2}{*}{$XDISIM1$}&
\multirow{2}{*}{1}&
{lots of equivalent deformation corresponding to generators redefinition}\\
&($SIM$ deformed)& & & additional accompanied dilatation for rotation and boost operation\\
\cline{2-5}
&
$XDISIM2$&
\multirow{2}{*}{$XDISIM2$}&
\multirow{2}{*}{1}&
{additional accompanied dilatation for rotation operation}\\
&($SIM$ deformed)& & & additional accompanied dilatation for boost operation\\
\hline

\multirow{5}{*}{$IHOM$} &
{$DIHOM1$}&
$DIHOM1$&
\multirow{3}{*}{1}&
{lots of equivalent representations corresponding to generators redefinition}\\
&($WDISIM$ )& ($WDISIM$ )& & additional accompanied dilatation for  boost operation\\
& & & & same structure as the corresponding part of $DISIM$\\
\cline{2-5}
&
$DIHOM2$&
$DIHOM2$&
\multirow{2}{*}{1}&
no natural representations inherited from Poincar\'e group\\
&($DIHOM$)&($DIHOM$) & & additional accompanied dilatation for  boost operation\\

\hline

\multirow{7}{*}{$TE(2)$} &
\multirow{2}{*}{$DTE1$}&
\multirow{2}{*}{$DTE1$}
&
\multirow{2}{*}{1}&
additional accompanied dilatation for rotation operation\\
 & & & &rotation operation not only in $xy$ plane but also in rotated $tz$ plane\\
\cline{2-5}
&\multirow{2}{*}{$DTE2$}&
{$DTE2a$}&
2 & translations are entangled with $t_1$ and $t_2$ operations\\
\cline{3-5}
& & $DTE2b$& 0 & no natural representation inherited from Poincar\'e group\\
\cline{2-5}
&
\multirow{3}{*}{$DTE3$}&
$DTE3a$&
2&
translations are entangled with $t_1$ and $t_2$ operations\\
\cline{3-5}
& &\multirow{2}{*}{$DTE3b$} &\multirow{2}{*}{1} & translations are entangled with $t_1$ and $t_2$ operations\\
& & & & rotation operation not only in $xy$ plane but also in rotated $tz$ plane\\
\hline

\multirow{4}{*}{$ISO(3)$} &
\multirow{2}{*}{$DISO(3)1$}&
\multirow{2}{*}{$DISO(3)1$}&
\multirow{2}{*}{1}&
inequivalent representation corresponding to different sign of deform parameter\\
 & & & & only translations operations deformed\\
\cline{2-5}
&
\multirow{2}{*}{$DISO(3)2$}&
\multirow{2}{*}{$DISO(3)2$}
&
\multirow{2}{*}{3}&
three inequivalent representations \\
& & & & only translations operations deformed\\
\hline
\multirow{4}{*}{$ISO(2,1)$} &
\multirow{2}{*}{$DISO(2,1)1$}&
\multirow{2}{*}{$DISO(2,1)1$}&
\multirow{2}{*}{1}&
inequivalent representation corresponding to different sign of deform parameter\\
 & & & & only translations operations deformed\\
\cline{2-5}
&
\multirow{2}{*}{$DISO(2,1)2$}&
\multirow{2}{*}{$DISO(2,1)2$}
&
\multirow{2}{*}{2}&
two inequivalent representations \\
& & & & only translations operations deformed\\

\hline
\hline



\end{tabular}
\end{center}
\end{table}

In summary, the deformation of Poincar\'e group itself is the de Sitter group which is the isometry of maximal symmetric space of four dimensional spacetime, i.e. the isometry group of a curved de Sitter spacetime.

The deformation of $ISIM$ can be classified into two families. One family is $DISIM$, in which $SIM$ part is undeformed. There are many equivalent deformations which are connected with each other by redefinition of generators. For some case there are a family of equivalent natural representations. The rotation and boost operation obtain additional accompanied scale transformation in all cases. The other family, in which the $SIM$ part is deformed, can be divided into two subfamilies. The first subfamily is $XDISIM1$. Similar to family $DISIM$, there are also many equivalent deformations which are connected with each other by redefinition of generators. There are also a family of equivalent natural representations. Both deformed $R_z$ and deformed $B_Z$ obtain additional accompanied scale transformation. The second subfamily  $XDISIM1$ also has a family of equivalent natural representations and both deformed $R_z$ and deformed $B_Z$ obtain additional accompanied scale transformation. The deformed rotation operation can be a meaningful rotation only if the additional accompanied scale factor is one, i.e, the corresponding deform parameter vanishes.

The deformation of $IHOM$ with $HOM$ part undeformed can be classified into two families. The first family is $DIHOM1$ which is the same deformed group $XDISIM1$ in lack of one generator $r_z$. The natural representation is the same as $XDISIM1$. The other family $DIHOM2$ is totally different from $DIHOM1$. The 5-d representation of $DIHOM2$ reveals that it is not the natural representation inherited from Poincar\'e group. The $DIHOM2$ should be the symmetry group of a curved spacetime similar to de Sitter group.

The deformation of $TE$ with $E(2)$ part undeformed can be classified into three families. In the first family $DTE1$, deformed rotation $R_z$ is not only a rotation in $xy$ plane but also a rotation in the rotated $tz$ plane and obtains additional accompanied scale transformation. The second family $DTE2$ can be further divide into two subfamilies. For the first subfamily $DTE2a$, there are two inequivalent natural representations in which only the translation operators are deformed and the deformed translation operators are translation entangled with $t_1$ and $t_2$. The second subfamily $DTE2b$ does not have a natural representation. Like $DTE2$, the third family $DTE3$ has two subfamilies. Just like $DTE2a$, the first one $DTE3a$ has two inequivalent natural representations in which only the translation operators are deformed and the deformed translation operators are translation entangled with $t_1$ and $t_2$. In the second subfamily $DTE3b$, the deformed translation operators are translation entangled with $t_1$ and $t_2$ and the deformed rotation $R_z$ is not only a rotation in $xy$ plane but also a rotation in the rotated $tz$ plane without additional accompanied scale transformation.

The deformation of $ISO(3)$ with $SO(3)$ part undeformed can be classified into two families. In the first family $DISO(3)1$, there are two inequivalent natural representations which correspond the sign of the deform parameter and only the translation operators deform. In the second family $DISO(3)2$, there are three inequivalent natural representations in which still only the translation operators deform.

Very similar to the case of $ISO(3)$, the deformation of $ISO(2,1)$ with $SO(2,1)$ part undeformed can be classified into two families. The first family $DISO(2,1)1$ is similar to the case of $DISO(3)1$ while the difference is that $DISO(2,1)2$ has two inequivalent natural representations.

With these detailed representation and deformed as well as undeformed operators' formalism, one can search the geometry which metric function is invariant under the action of the specified semi-product POincar\'e subgroup and its deformed partner and then construct the field theory in spacetime which the invariant metric function corresponds to. This procedure will build up the field theory realization of Cohen-Glashow's proposal of VSR. In our subsequent work we will present the search of invariant metric function and the construction of field theory.

\bigskip

\acknowledgments{}
X.Xue wish to thank Hanqing Zheng for illuminating discussions.


\bibliographystyle{unsrt}

\end{document}